\documentclass[12pt]{article}
\usepackage{times}
\usepackage{epsfig}
\usepackage{amsmath}
\usepackage{amssymb}
\usepackage{amsthm}
\usepackage{epstopdf}
\usepackage{graphicx}
\usepackage{enumerate}
\usepackage{authblk}
\usepackage{array}
\usepackage{cite}
\usepackage{color}
\usepackage[margin=1.in]{geometry}
\usepackage[caption=false]{subfig}
\usepackage{float}
\normalsize \baselineskip 16pt
\newcommand {\beq} {\begin{equation}}
\newcommand {\eeq} {\end{equation}}

\newcolumntype{P}[1]{>{\centering\arraybackslash}p{#1}}
\newcolumntype{M}[1]{>{\centering\arraybackslash}m{#1}}

\begin{document}
\title{Moving discrete breathers in a $\beta$-FPU lattice revisited}
\author[1]{Henry Duran}
\author[2,3]{Jes\'us Cuevas--Maraver}
\author[4]{Panayotis\ G.\ Kevrekidis}
\author[1]{Anna Vainchtein}
\affil[1]{\small Department of Mathematics, University of Pittsburgh, Pittsburgh, Pennsylvania 15260, USA}
\affil[2]{\small Grupo de F\'{\i}sica No Lineal. Departamento de F\'{\i}sica Aplicada I, Escuela Polit\'{e}cnica Superior, Universidad de Sevilla. C/Virgen de \'Africa, 7, Sevilla 41011, Spain}
\affil[3]{\small Instituto de Matem\'{a}ticas de la Universidad de Sevilla (IMUS). Edificio Celestino Mutis, Avda. Reina Mercedes s/n, 41012-Sevilla, Spain}
\affil[4]{\small Department of Mathematics and Statistics, University of
Massachusetts, Amherst, MA 01003-9305, USA}

\maketitle

\begin{abstract}
  In the present work we revisit the existence, stability and dynamical properties of moving discrete breathers
  in $\beta$-FPU lattices. On the existence side, we propose a numerical procedure, based on a continuation
  along a sequence of velocities, that allows to systematically construct breathers
  traveling more than one lattice site per period. On the stability side, we explore the stability spectrum of the obtained
  waveforms via Floquet analysis
and connect it to the energy-frequency bifurcation diagrams.
We illustrate in this context examples of the
  energy being a multivalued function of the frequency, showcasing the coexistence of different moving
  breathers at the same frequency. Finally, we probe the moving breather dynamics and observe how
  the associated instabilities change their speed, typically slowing them down over long-time simulations.

\end{abstract}

\section{Introduction}
Discrete breathers are time-periodic nonlinear modes that arise in lattices due to the interplay of dispersion and nonlinearity \cite{Aubry97,Aubry06,Flach08}. The most common form of such excitations are stationary bright breathers, originally called intrinsic localized modes \cite{SievTak88,Page90,SandPageSchm92} due to their spatial localization. Breathers were found to exist in Hamiltonian and damped-driven lattices and were experimentally observed in a variety of nonlinear discrete systems, including Josephson junction arrays \cite{Binder00,Trias00}, forced-damped arrays of coupled pendula \cite{Cuevas09}, electrical lattices \cite{Palmero11,English13,remoissenet}, 
micromechanical systems \cite{Sato03,Sato04,Sato06}, the denaturation
of the DNA double strand~\cite{Peybi} and granular chains \cite{Boechler10,Chong14,Zhang17,granularBook}.

In the years since the breathers were first discovered \cite{Ovchinnikov70}, there has been much progress in understanding their existence, spectral stability and dynamical properties~\cite{Aubry06,Flach08,Kevrekidis11}.
For instance, explicit criteria for linear~\cite{jcmprl} and nonlinear~\cite{pelin2} stability of stationary breathers have been put forth.
Additionally, it was observed that instability of stationary breathers sets them in motion, and long-lived traveling breathers have been found numerically in various nonlinear lattices \cite{Burlakov90,Hori92,SandPageSchm92,ChenAubTsi96,Cretegny98}.
Breather mobility is of considerable interest because it is associated with energy transport in the lattice; indeed
such coherent structures have been proposed as a prototypical means for achieving targeted
energy transfer in discrete nonlinear systems~\cite{kopid}. An exact moving breather is time-periodic modulo a shift by one or more lattice spaces. The period is an integer multiple of the period of internal vibrations. Such solutions have been constructed using the Newton iterative method, e.g., for Klein-Gordon \cite{Cretegny98,Aubry98,Archilla19} 
 and $\beta$-FPU
 \cite{Cretegny98,YoshiDoi07} lattices. For generic interaction potentials that do not possess a certain symmetry \cite{doi16,doi20}, moving breathers are no longer spatially localized: instead, they possess oscillatory wings whose amplitude depends on the internal breather frequency and its propagation velocity.

The first detailed analysis of this dependence for a $\beta$-FPU
lattice was performed in \cite{YoshiDoi07}. The authors constructed
numerically exact moving breathers for several different rational
values of the \emph{period-wise}
velocity $V_1=r/s$, where $r$ is the number of lattice sites the breather travels over $s$ periods of the internal vibration. Performing a continuation in internal frequency $\omega$ at fixed $V_1$, they investigated how the wing energy (or, equivalently, amplitude) of these breathers depends on their internal frequency. In particular, they studied the mechanism for resonances in the wing amplitude and derived an approximate formula for the resonant frequencies. They also briefly summarized the results of linear stability investigation (without providing a systematic analysis thereof) for the computed solutions.

Motivated by these earlier studies, we revisit the problem and conduct a more detailed investigation of the breather existence, stability, dynamics and resonance features.
Our analysis extends the results of the earlier work in several ways.
Importantly, we consider moving breathers propagating by more than one lattice space ($r>1$) over its period, extending the earlier work that had focused chiefly on the $r=1$ case. To compute such solutions, we developed a numerical procedure based on a continuation along a sequence of rational values of $V_1$.
We show that the total breather energy (the Hamiltonian) and the wing energy are in fact  \emph{multivalued} functions of the internal frequency $\omega$, so that there are several moving breathers with the same $\omega$ and different energies. Moreover, our results reveal the truly nonlinear form of the resonances: a rapid increase in wing energy is followed by a more gradual one. Subsequently, we provide a detailed analysis of the linear stability of the obtained solutions focusing on the consequences of the instability associated with real Floquet multipliers $\mu>1$. In particular, we investigate the dynamics of the breathers perturbed along the corresponding unstable eigenmodes and show that after repeated interactions with the wing oscillations due to the periodic boundary conditions the breather gradually decelerates and eventually becomes nearly stationary, with its velocity oscillating around zero.

The paper is organized as follows. We formulate the problem in Sec.~\ref{sec:formulation} and describe our numerical procedures in Sec.~\ref{sec:methods}. In Sec.~\ref{sec:freq} we examine the dependence of the moving breather with different period-wise
velocities on the internal frequency and discuss the multivalued nature of the obtained energies, resonances and stability. Consequences of the observed real instabilities in the breather dynamics are explored in Sec.~\ref{sec:instab}. Concluding remarks can be found in Sec.~\ref{sec:conclusions}, along with some suggestions for future work. In the Appendix, we discuss additional solutions that coexist with the ones described in the main text but have different linear spectra.

\section{Problem formulation}
\label{sec:formulation}
We consider a lattice of $N$ particles with nearest-neighbor interactions governed by a $\beta$-FPU potential. In dimensionless variables the Hamiltonian of the system is given by
\beq
H=\frac{1}{2}\sum_{n=1}^N p_n^2+\sum_{n=1}^N\left(\frac{1}{2}(q_{n+1}-q_n)^2+\frac{\beta}{4}(q_{n+1}-q_n)^4\right)=\sum_{n=1}^N e_n,
\label{eq:H}
\eeq
where $q_n$ denotes the displacement of the $n$th particle, $p_n=\dot{q}_n=dq_n/dt$ is its momentum (the mass is rescaled to unity), $\beta$ measures the strength
of the nonlinear coupling, and
\beq
e_n=\frac{1}{2}p_n^2+\frac{1}{4}\left[(q_{n+1}-q_n)^2+(q_n-q_{n-1})^2\right]+\frac{\beta}{8}\left[(q_{n+1}-q_n)^4+(q_n-q_{n-1})^4\right]
\label{eq:en}
\eeq
is the site energy density. The equations of motion are
\beq
\ddot{q}_n=q_{n+1}+q_{n-1}-2q_n+\beta\left[(q_{n+1}-q_n)^3-(q_n-q_{n-1})^3\right].
\label{eq:EoM}
\eeq
In what follows, we assume that $N$ is even and prescribe periodic boundary conditions: $q_{n+N}=q_n$, $p_{n+N}=p_n$. In the numerical results presented in this work we set $\beta=1$.

The $\beta$-FPU problem \eqref{eq:EoM} is known to have two types of stationary discrete breather solutions $q_n(t)=x_n(t)$ that are time-periodic, $x_n(T)=x_n(0)$, and spatially localized in terms of the relative displacements $x_n-x_{n-1}$. Here, $T=2\pi/\omega$ is the period of internal oscillations with frequency $\omega$. The first type is the site-centered Sievers-Takeno (ST) mode \cite{SievTak88}, with displacement that has even symmetry about the center, and the second type is the bond-centered Page (P) mode \cite{Page90}, with odd displacement.
The P mode is linearly stable, while the ST mode is unstable \cite{SandPageSchm92}. Perturbing an ST mode along an eigenmode corresponding to the instability sets the breather in motion.

Our focus here is on {\it moving} discrete breathers that propagate $r$ lattice sites over $s$ periods $T=2\pi/\omega$ of internal oscillations and satisfy \cite{YoshiDoi07}
\beq
\left[\begin{array}{c}
  \{q_n(sT)\}_{i=1}^N \\ \{p_n(sT)\}_{n=1}^N \\  \end{array}\right]
- (-1)^r\left[\begin{array}{c}
 \{q_{n-r}(0)\}_{n=1}^N \\ \{p_{n-r}(0)\}_{n=1}^N \\  \end{array}\right]=\mathbf{0},
\label{eq:MovCond}
\eeq
where the indices are mod $N$ due to periodic boundary conditions.
Here $s$ and $r$ are integers, and
\beq
V_1=\dfrac{r}{s}
\label{DoiVel}
\eeq
denotes the period-wise velocity of the breather (the number of lattice sites transversed over the period of one internal oscillation), while its translational velocity is given by
\beq
V_2=\dfrac{V_1}{T}=\dfrac{r}{sT}.
\label{eq:ArcVel}
\eeq

\section{Numerical Methods}
\label{sec:methods}
To obtain moving breathers, we must find fixed points of the map defined by \eqref{eq:MovCond} using the Newton iterative method, with an appropriately perturbed unstable ST stationary breather, whose instability induces the breather mobility, as an initial seed. Here and in what follows, we use symplectic and explicit fourth-order Runge-Kutta-Nystr\"om algorithm \cite{calvo93} to integrate the equations of motion. We found that over the course of the simulations, the maximum absolute relative difference of the total energy when compared to the initial total energy is on the order of $10^{-10}$.
We start by constructing an ST breather $x_n(t)$ with the given internal frequency $\omega$, using the Newton iterative method and numerical continuation from the anticontinuous limit \cite{Marin96}. Linearizing \eqref{eq:EoM} around the ST breather by setting $q_n(t)=x_n(t)+\epsilon y_n(t)$ and considering $O(\epsilon)$ terms, we obtain
\[
\ddot{y}_n-(y_{n+1}+y_{n-1}-2y_n)-3\left((x_{n+1}-x_n)^2(y_{n+1}-y_n)-(x_n-x_{n-1})^2(y_n-y_{n-1})\right)=0,
\]
which is used to compute the monodromy matrix $\mathcal{F}$ defined by
\beq
\begin{bmatrix}
\mathbf{y}(T) \\
\dot{\mathbf{y}}(T)
\end{bmatrix}
=
\mathcal{F}
\begin{bmatrix}
\mathbf{y}(0) \\
\dot{\mathbf{y}}(0)
\end{bmatrix},
\eeq
where the vector functions $\mathbf{y}(t)$ and $\dot{\mathbf{y}}(t)$ have components $y_n(t)$ and $\dot{y}_n(t)$, respectively.
The Floquet multipliers $\mu$ are obtained by finding the eigenvalues of $\mathcal{F}$, once the
iterative procedure has converged.
A Floquet multiplier satisfying $\vert \mu \vert >1$ indicates instability. An ST mode has an unstable eigenmode corresponding to a real Floquet multiplier $\mu>1$; naturally, due to the Hamiltonian nature of the problem,
there exists a complementary (inverse) one with $\mu<1$. Following \cite{ChenAubTsi96}, we obtain the initial seed for a moving breather by applying a kinetic perturbation of the ST breather. Specifically, we use the momentum part $\mathbf{\delta p}$ of the eigenvector associated with the instability, so that our initial guess for the moving breather is given by
\beq
\begin{bmatrix}
\mathbf{q} \\
\mathbf{p}
\end{bmatrix}
=
\begin{bmatrix}
\mathbf{x}(0) \\
\mathbf{0}
\end{bmatrix}
+
\lambda
\begin{bmatrix}
\mathbf{0} \\
\mathbf{\delta p}
\end{bmatrix},
\label{eq:guess}
\eeq
where $\lambda$ is the strength of the perturbation.

To construct moving breathers with $V_1=1/s$ for some integer $s \geq 1$, we use the Newton iterative method to find fixed points of \eqref{eq:MovCond} with initial guess \eqref{eq:guess} as the values of $\lambda$ are being incremented within some interval. We typically start with $\lambda=-1$ and increase it by $10^{-2}$ up to $\lambda=1$. Once this has been completed, we look at the solutions for which the square of the $\ell^2$ norm of the objective function of
the Newton iteration, defined by the left hand side of \eqref{eq:MovCond},
is below some threshold. Doing this allows us to obtain moving
breathers on different branches in the ($\omega$, $H$) plane near the
resonance values of $\omega$, as described below. Solutions with other
frequency values are then found using parameter continuation along each branch. Typically, this continuation was done in $\omega$, but near the turning points for $\omega$ we used $H$ as a continuation parameter.
We found that this method successfully generates moving breathers with $r=1$ but has not worked in the examples we considered for velocities with $r>1$.

\begin{figure}[!htb]
\centering
\subfloat[]{\includegraphics[width=0.5\textwidth]{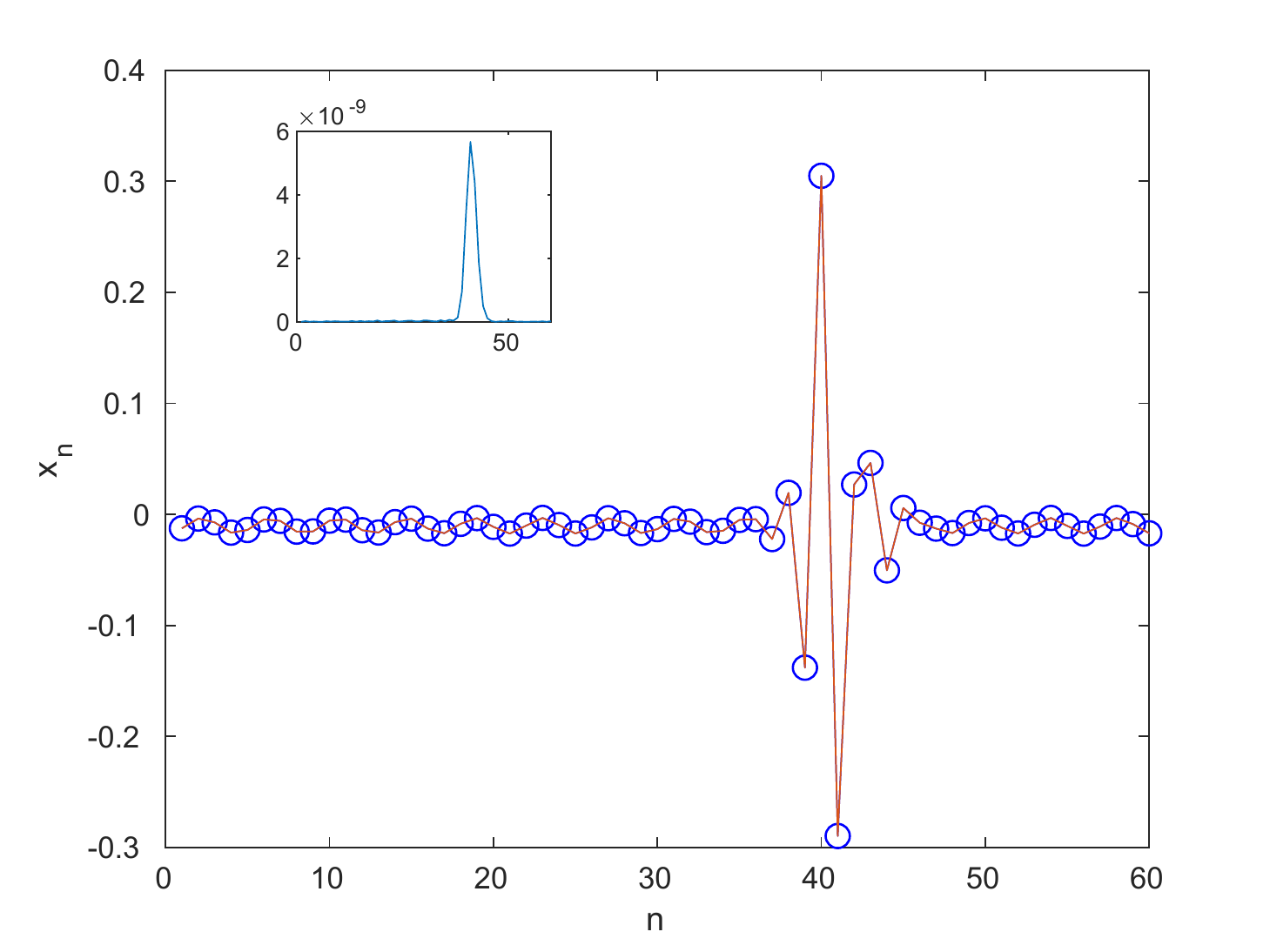}}
\subfloat[]{\includegraphics[width=0.5\textwidth]{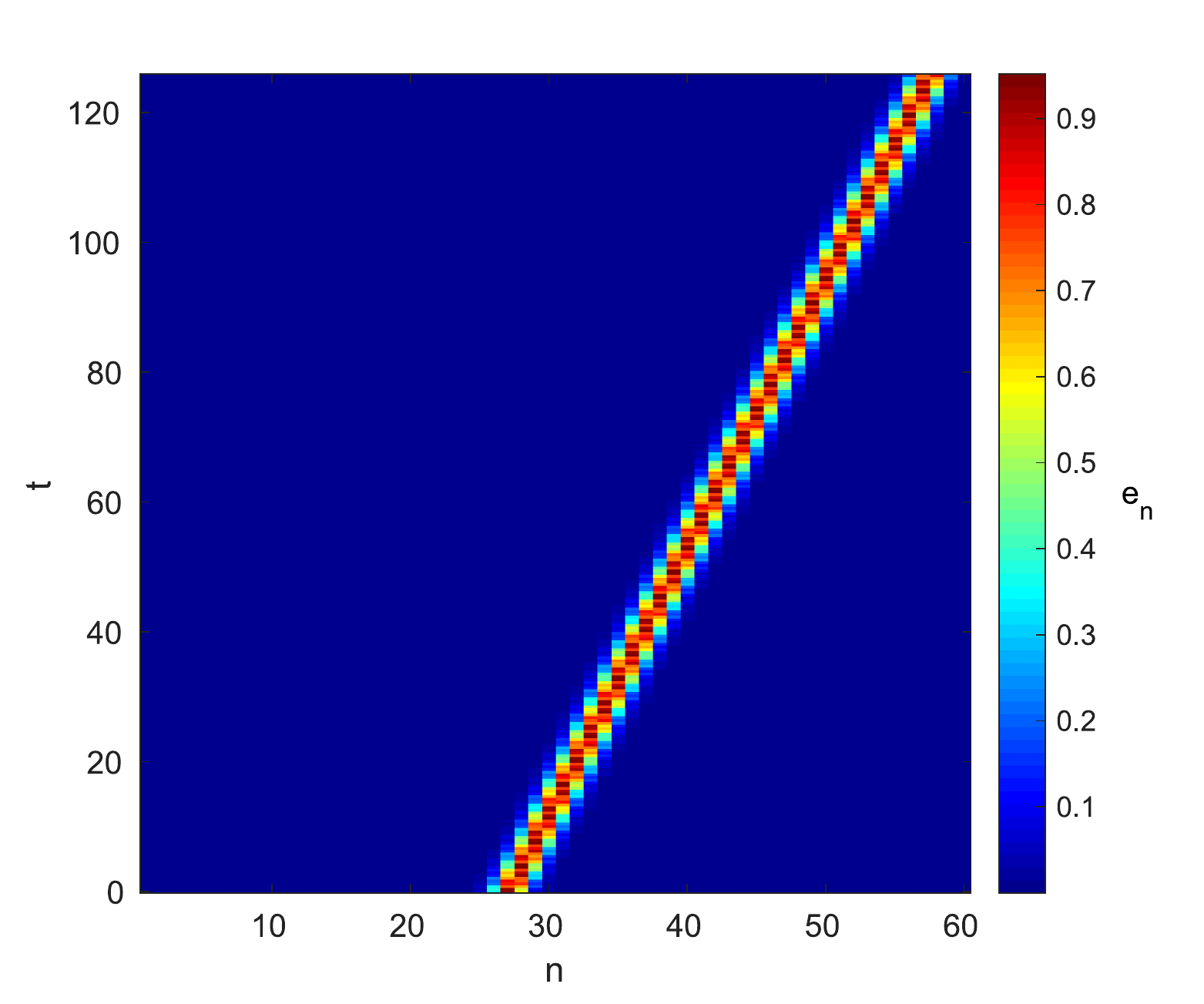}}\\
\subfloat[]{\includegraphics[width=0.5\textwidth]{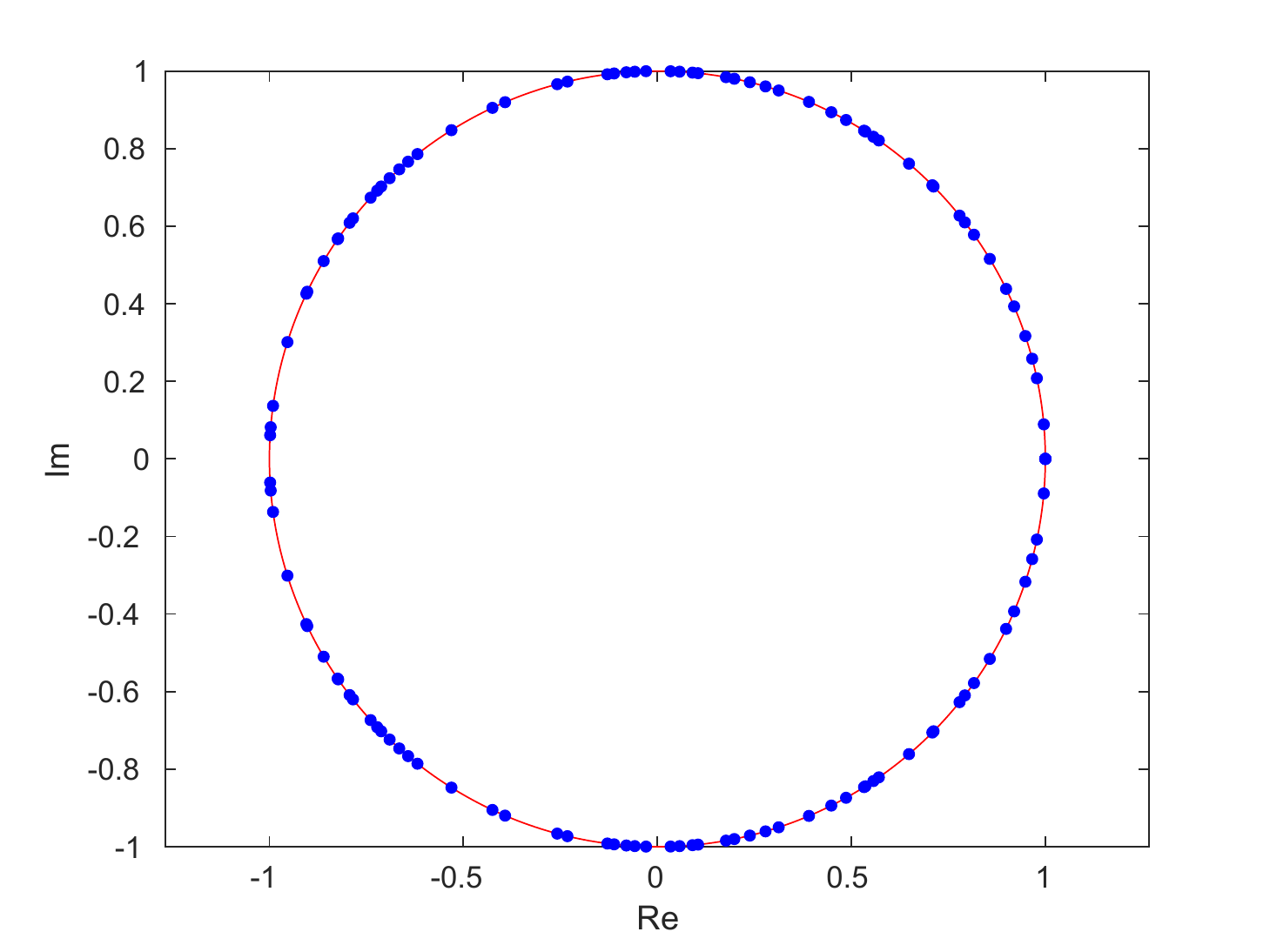}}
\caption{\footnotesize (a) Verification of the relation described in \eqref{eq:MovCond} for the moving breather with period-wise velocity $V_1=14/23$ and internal frequency $\omega=2.5$. The blue circles are the displacements at time $t=23T$, while the solid red line is the displacement at time $t=0$ shifted to the right by $14$ lattice sites. The inset shows the absolute difference between the two sets of displacements.  (b) Space-time evolution of the site energy $e_n(t)$. (c) Floquet multipliers $\mu$ associated with the linearization around the solution of panels (a)-(b). The absence
of multipliers lying off of the unit circle suggests the spectral stability of the relevant waveform.}
\label{fig:v14_23}
\end{figure}

To compute moving breathers with period-wise velocity $V_1=r/s$, where $r>1$, we have developed  the following numerical procedure. We use one of the moving breathers with $V_1=1/s_0$ for some integer $s_0$ as an initial guess and construct a monotone sequence $v_1,v_2,...,v_k$ of rational values of the period-wise velocity that are close enough together and satisfy $v_1=1/s_0$ and $v_k=r/s$. These values are chosen in a way that minimizes $s$ while staying within a prescribed step difference, empirically selected to be  between $0.018$ and $0.022$. Depending on the value of $\omega$ chosen, it is possible that for one of the chosen $v_i$, the moving breather solution will be close to a resonance; in this case, a larger step in $v_i$ is needed to bypass the resonance. For example, to compute a moving breather with $V_1=5/7$ we used the sequence
\[
\lbrace v_{i}\rbrace = \lbrace 1/2,\, 12/23,\, 13/24,\, 9/16,\, 7/12,\, 23/38,\, 5/8,\, 20/31,\, 2/3,\, 11/16,\, 5/7 \rbrace.
\]
We then use a continuation procedure that involves obtaining the moving breather with velocity $v_i$ using Newton's iterative method and the breather with velocity $v_{i-1}$ as the initial guess. An example of a moving breather with $V_1=14/23$ and $\omega=2.5$ obtained using this method is shown in Fig.~\ref{fig:v14_23}(a,b).

To determine the stability of a computed moving breather, we linearize \eqref{eq:EoM} about it and construct the monodromy matrix $\mathcal{F}$ defined by
\beq
\begin{bmatrix}
\{y_{n+\tilde{r}}(\tilde{s}T)\}_{n=1}^N \\
\{\dot{y}_{n+\tilde{r}}(\tilde{s}T)\}_{n=1}^N
\end{bmatrix}
=
\mathcal{F}
\begin{bmatrix}
\{y_n(0)\}_{n=1}^N \\
\{\dot{y}_n(0)\}_{n=1}^N
\end{bmatrix},
\label{eq:MonoMovFix}
\eeq
where $\tilde{r}=r$, $\tilde{s}=s$ if $r$ is even and $\tilde{r}=2r$, $\tilde{s}=2s$ if $r$ is odd. Fig.~\ref{fig:v14_23}(c) shows the Floquet multipliers for the breather with $V_1=14/23$ and internal frequency $\omega=2.5$. This breather appears to be linearly stable.

To explore the consequences of an instability associated with a real Floquet multiplier $\mu>1$ for a moving breather, we perturb it along the corresponding eigenmode and use the following method to approximate the translational velocity $V_2$ of the ensuing waveform as a function of time. The procedure involves computing the location of the center of the energy density of the moving breather. We divide the time interval $[t_i,t_f]$, where $t_i$ is the initial and $t_f$ is the final time, into subintervals of length $\Delta t$, thus selecting sample times $t_i$ such that $t_{j+1}-t_j = \Delta t$. Typically, we set $\Delta t = sT$, where $T$ is the internal period and $s$ is the number of periods the unperturbed breather needs to advance $r$ sites. At each time $t_j$, we compute the energy density $e_{n,j}$ and use it to obtain an approximation for the center $X_j$ of the wave
\beq
X_j = \frac{\sum_{n \in \{{\rm core}\}} {ne_{n,j}}}{\sum_{n \in \{{\rm core}\}}{e_{n,j}}}.
\label{eq:weightedsum}
\eeq
In order to improve the accuracy of this approximation, we use a spline interpolation of the energy density. We then compute \eqref{eq:weightedsum} including the interpolated points in the core of the moving breather. To determine the width of the core, we start from the maximum of the energy density. We then traverse the chain until the absolute difference between the energy density and wing energy, which is determined by averaging the ten particles that make up the wings, is on the order of $10^{-4}$. The distance between the particle where the maximum occurs and the cutoff particle is half of the core width.
We choose as a center point the maximum of the interpolated energy density. Once the weighted energy center has been found, we repeat the above procedure using the weighted energy center as the center point. This has little effect for waveforms with small-amplitude wings, but when the wings have larger amplitude, the recalculation is necessary to compensate for the effect they have on the energy density as the center crosses a boundary. The translational velocity $V_2(t)$ of the wave is then approximated by
\beq
V_2(t_j) \approx \frac{X_{j+1}-X_j}{t_{j+1}-t_j}.
\eeq

\section{Frequency dependence, resonances and stability}
\label{sec:freq}
\begin{figure}[!htb]
\centering
\subfloat[]
{\includegraphics[width=0.5\textwidth]{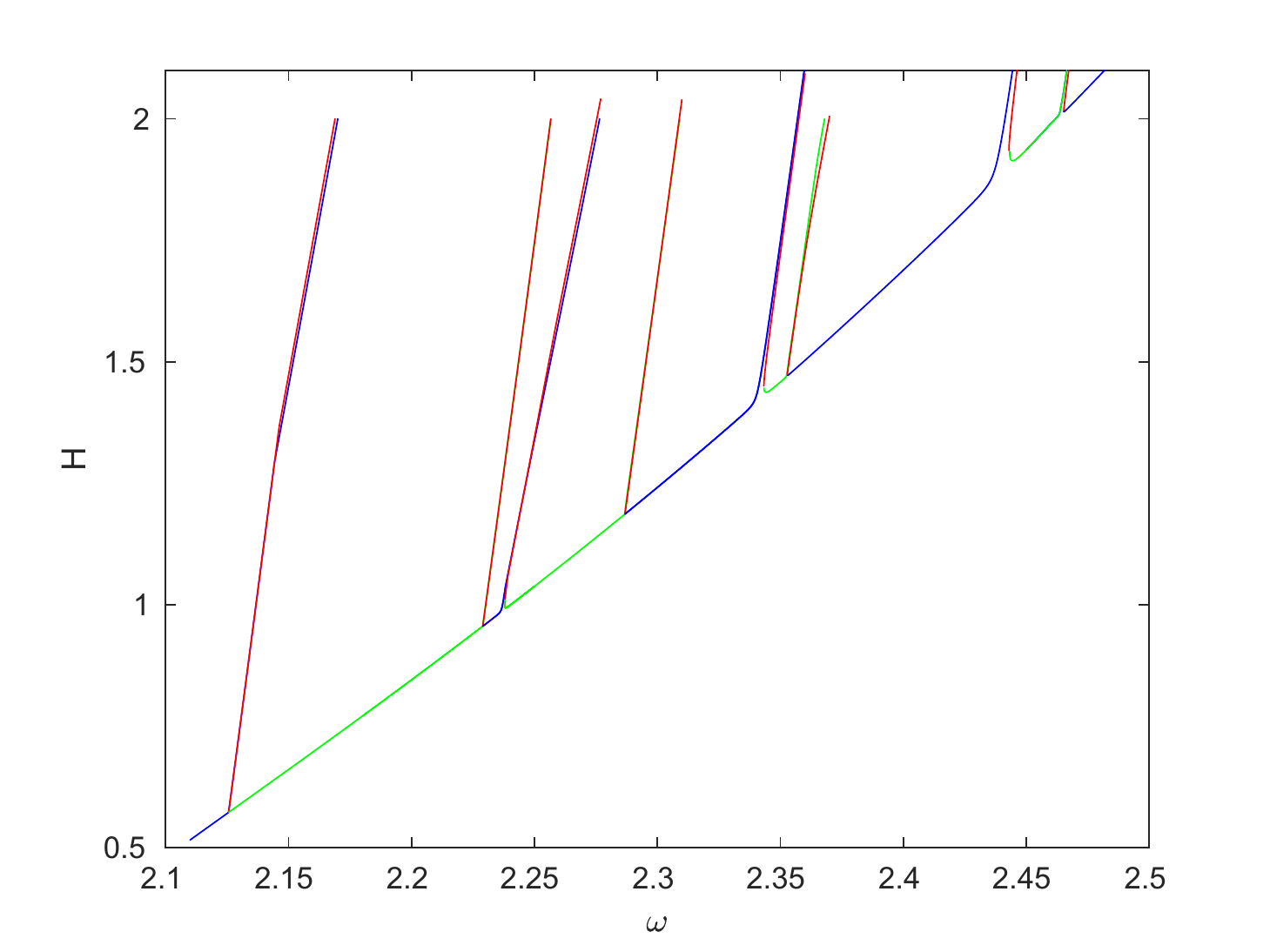}}\\
\subfloat[]
{\includegraphics[width=0.5\textwidth]{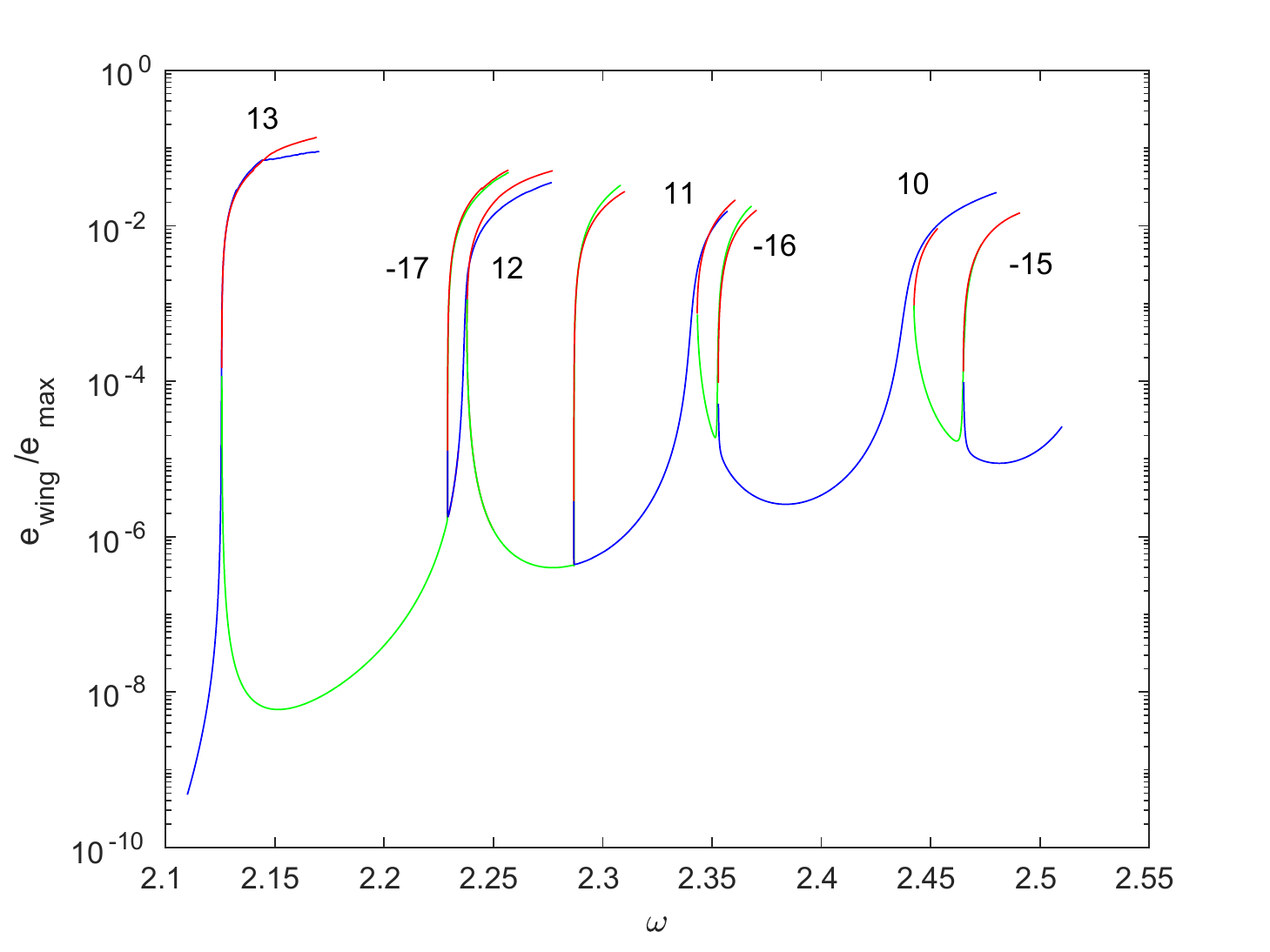}}
\subfloat[]
{\includegraphics[width=0.5\textwidth]{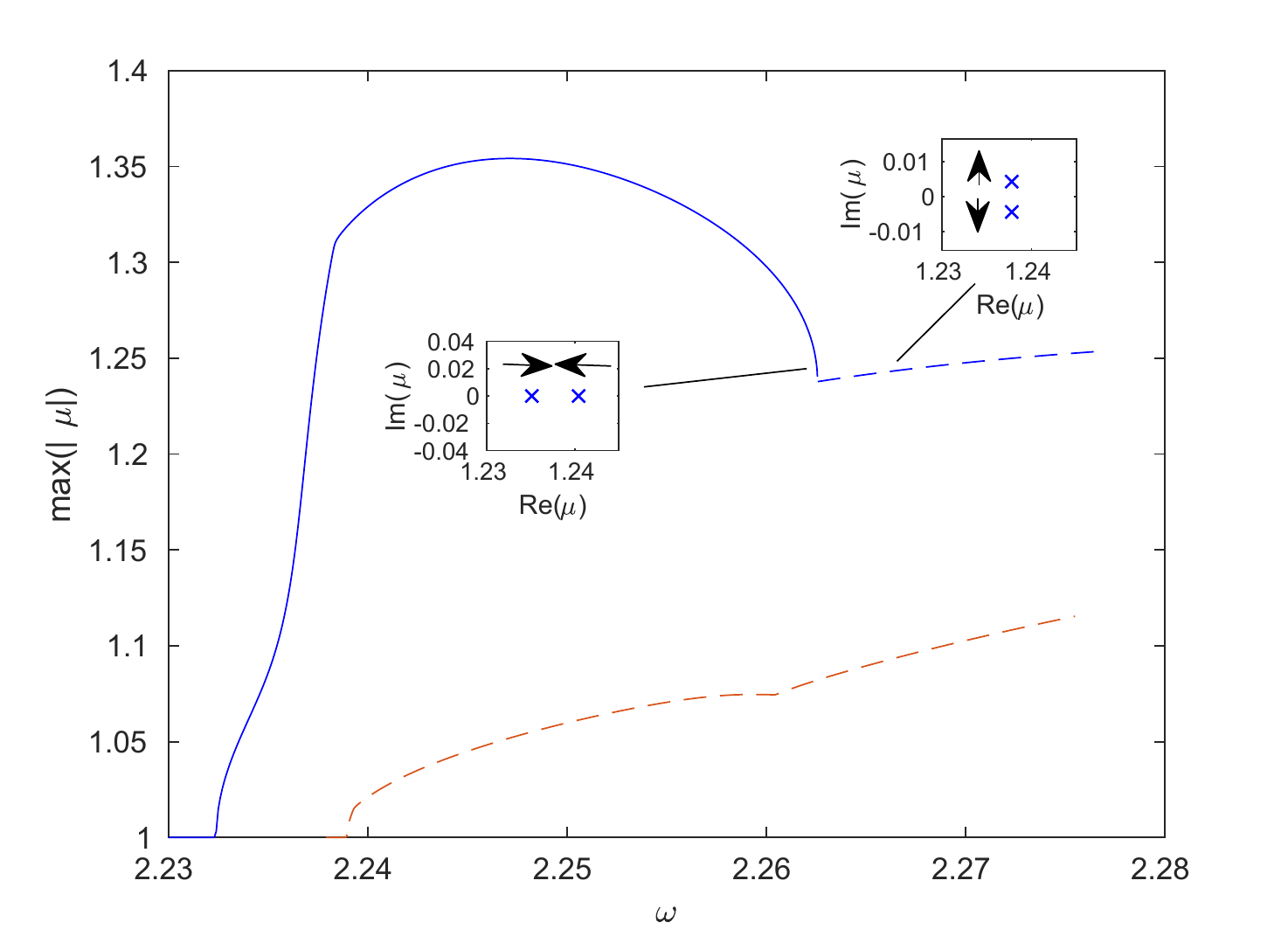}}
\caption{\footnotesize (a) Energy $H$, (b) the normalized average site energy $e_{wing}/e_{max}$ of the wings and (c) the maximum moduli of the Floquet multipliers $\mu$ along different branches as functions of $\omega$ at $V_1 = 1/3$ and $N=60$ near the resonance $\omega = 2.237$. The Floquet multiplier with the maximum modulus has nonzero real and imaginary parts along dashed portions the curve and is real along the solid one. Insets illustrate that this transition occurs due to the collision of a pair of real Floquet multipliers and subsequent emergence of a quadruplet of complex-valued multipliers symmetric about the unit circle (only the pair of such multipliers outside the unit circle is shown in the second inset). Different colors correspond to different branches in (a). The numbers in (b) are the values of $m$ for each resonance (see the text for detail). }
\label{fig:m60v1_3_full}
\end{figure}

\begin{figure}[!htb]
\centering
\includegraphics[width=0.5\textwidth]{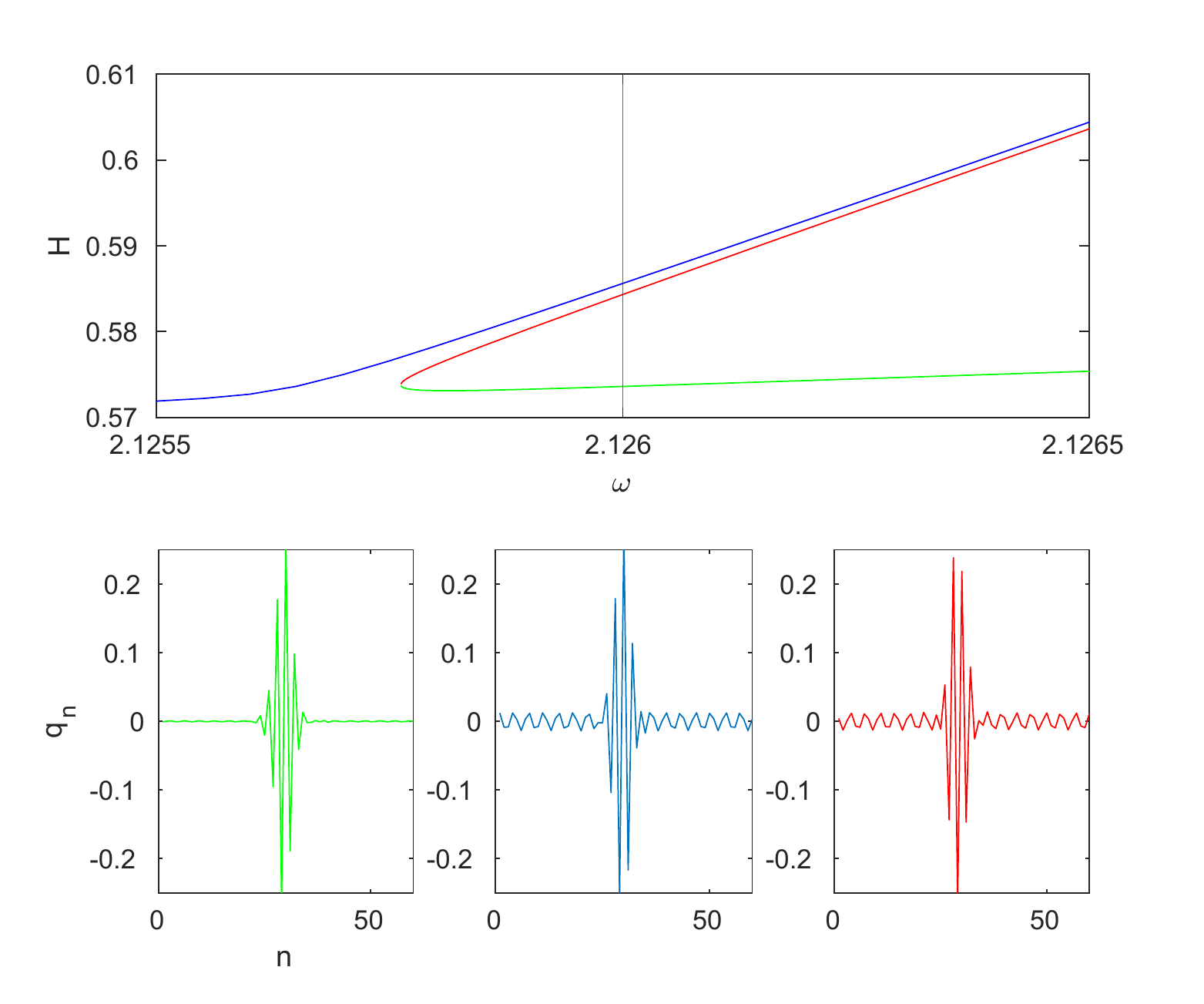}
\caption{\footnotesize The top panel shows a zoomed-in view of the resonance near $\omega = 2.126$. The black vertical line marks $\omega = 2.126$ at which three different moving breathers coexist. These breathers are shown in the three bottom panels, where colors match the respective branches depicted in the top panel. Here $V_1 = 1/3$ and $N = 60$.}
\label{fig:m60v1_3_multivalued}
\end{figure}

We now investigate the dependence of the moving breather solutions on the internal frequency $\omega$ at fixed period-wise velocity $V_1$ and the lattice size $N$.
The results for $V_1=1/3$ and $N=60$ are shown in Fig.~\ref{fig:m60v1_3_full}.
Panel (a) shows the total energy (Hamiltonian) $H$ as a function of $\omega$. One can see that there is a number of resonances at certain frequency values. At these values, the amplitude of the wing oscillations rapidly increases. Near the resonance frequencies, the breather energy is a \emph{multivalued} function of $\omega$. Indeed, near each resonance frequency, the curve can be split into three pieces: the top branch, the middle branch, and the bottom branch, where the middle and bottom branches are connected to each other by a turning point.
We distinguish between the bottom and top branches by alternating colors between green and blue at each resonance. Note that the bottom branch corresponding to one resonance frequency eventually merges with the top branch near another resonance. Along the three branches near each resonance, there are distinct moving breathers with the same internal frequency, as illustrated in Fig.~\ref{fig:m60v1_3_multivalued} for $\omega = 2.126$.

Panel (b) of Fig.~\ref{fig:m60v1_3_full} shows the corresponding average site energy in a wing portion of the breather, normalized by the maximum site energy. One can clearly see the \emph{nonlinear} character of the resonances, with rapid increase in wing energy followed by a more gradual one. Both the nonlinear form of the resonances and the multivalued nature of the frequency dependence were, apparently, missed in the earlier computations \cite{Cretegny98,YoshiDoi07}.

Yoshimura and Doi in \cite{YoshiDoi07} used a normal mode analysis to approximate resonance frequencies. For completeness, we briefly describe the main steps of their derivation. The normal mode coordinates $Q_m(t)$, $m=-N/2-1,\dots,N/2$, are defined by \cite{poggi97}
\[
q_n(t)=\dfrac{(-1)^n}{\sqrt{N}}\sum_{m=-(N/2-1)}^{N/2} Q_m(t)\left[\cos\left(\dfrac{2\pi}{N}mn\right)-\sin\left(\dfrac{2\pi}{N}mn\right)\right], \quad n=1,2,\dots,N
\]
and have the natural frequency associated with the dispersion relation:
\beq
\Omega_m=2\cos\left(\dfrac{\pi m}{N}\right).
\label{eq:Omega_m}
\eeq
For stationary breathers these modes are approximated in \cite{YoshiDoi07} using the method in \cite{kosevich93} with frequency $\omega$, which yields $Q_m(t) \approx A_m\cos(\omega t)$, where
\beq
A_m = \frac{\pi}{2\sqrt{6\beta N}}\text{sech}\left[\frac{\pi^2m}{N\sqrt{\omega^2-4}}\right].
\label{eq:Am}
\eeq
Complex normal modes $U_m(t)=\frac{1}{2}(Q_m+Q_{-m})+\frac{i}{2}(Q_m-Q_{-m})$ are then used to
construct moving breather solutions with $V_1=r/s$ in the form
\[
U_m(t) = \psi_m(t)e^{-i \frac{mr}{Ns} \omega t},
\]
where $\psi_m(t)$ are complex-valued functions satisfying
\beq
\begin{split}
&\frac{d^2\psi_m}{dt^2} - i\frac{2mr\omega}{Ns}\frac{d\psi_m}{dt}+\left\{ \Omega_m^2 - {\left(\frac{mr\omega}{Ns}\right)}^2\right\}\psi_m\\
&= -\frac{\beta}{N}\sum_{i,j,k = -N_h}^{N_h}\Omega_m\Omega_i\Omega_j\Omega_k\psi_i\psi_j\psi_k
\cdot e^{i\left[\lbrace m-(i+j+k) \rbrace r/Ns\right]\omega t}\Delta(m-(i+j+k)),
\end{split}
\label{eq:psim_eqns}
\eeq
where $\Delta(m) = (-1)^m$ if $r=mN$ for $m \in \mathbb{Z}$ and zero otherwise, $N_h = N/2 -1$, and $\Omega_m$ is defined in \eqref{eq:Omega_m}. The solution of \eqref{eq:psim_eqns} is then sought in the form
\beq
\psi_m(t)=\psi_m^0(t)+u_m(t), \quad \psi_m^0(t) = \sum_{\ell = \pm 1} A_m e^{i\ell\omega t},
\quad u_m(t) = \sum_{n=-\infty}^{\infty} a_{m,n}e^{in(\omega/s)t},
\label{eq:psim_form}
\eeq
where $u_m(t)$ is the deviation from the stationary breather $\psi_m^0(t)$ with $A_m$ given by \eqref{eq:Am}, and both components are periodic functions with period $sT$ that are expanded in Fourier series, with coefficients $a_{m,n}$ for $u_m(t)$. Here, only the dominant fundamental frequency components are kept in the expansion for $\psi_m^0(t)=A_m\cos(\omega t)$. Substituting \eqref{eq:psim_form} into \eqref{eq:psim_eqns} and considering the leading-order approximation in terms of $u_m(t)$ results in a linear system for $a_{m,n}$. Analysis of this system shows that $\vert a_{m,n} \vert$ becomes large when its coefficient is close to zero. Setting these coefficients to zero thus yields an approximation for the resonance frequency values $\omega_m$, $\vert m \vert < N/2$,
at which the $m$th normal mode is excited. The approximate resonance condition \cite{YoshiDoi07} is given by
\beq
\left\vert \frac{n}{s} - \frac{mr}{Ns}\right\vert \omega_m = \Omega_m\sqrt{1 + \frac{2}{N}\sqrt{\omega_m^2-4}},
\label{eq:Resonance}
\eeq
where $n$ may take values $n = \pm s$ or $n = \pm (s \pm r)$, depending on the frequency interval and the value of $V_1=r/s$, and we also recall \eqref{eq:Omega_m}.
Using \eqref{eq:Resonance}, we computed the values of $m$ and $\omega_m$ for each resonance shown in Fig.~\ref{fig:m60v1_3_full}(b); the corresponding values of $m$ are shown in the plot. Table~\ref{table:FreqMult_v1_3} compares the predicted values of resonance frequencies with the numerical ones.
\begin{table}[!htb]
\centering
\begin{tabular}{ |M{2cm}|M{2.25cm}|M{1.25cm}|}
 \hline
 numer $\omega_m$ & approx $\omega_m$ & $m$ \\
 \hline
 $2.126$ & $2.129$ & $13$ \\
 \hline
 $2.237$ & $2.244$ & $12$ \\
 \hline
 $2.342$ & $2.352$ & $11$ \\
 \hline
 $2.440$ & $2.454$ & $10$ \\
 \hline
  $2.464$ & $2.483$ & $-15$ \\
 \hline
  $2.352$ & $2.364$ & $-16$ \\
 \hline
  $2.229$ & $2.236$ & $-17$ \\
 \hline
\end{tabular}
\caption{Comparison of numerical and approximate resonance values
  $\omega_m$ for $V_1 = 1/3$ and $N=60$. The approximate values are
  computed using \eqref{eq:Resonance}. The numerical values were
  computed by using the wing energy plots, such as
  Fig.~\ref{fig:m60v1_3_full}(b), and estimating the frequency at the
  center of the gap that separates branches corresponding to each
  resonance.
}
\label{table:FreqMult_v1_3}
\end{table}
%

\begin{figure}[!htb]
\centering
\subfloat[]
{\includegraphics[width=0.5\textwidth]{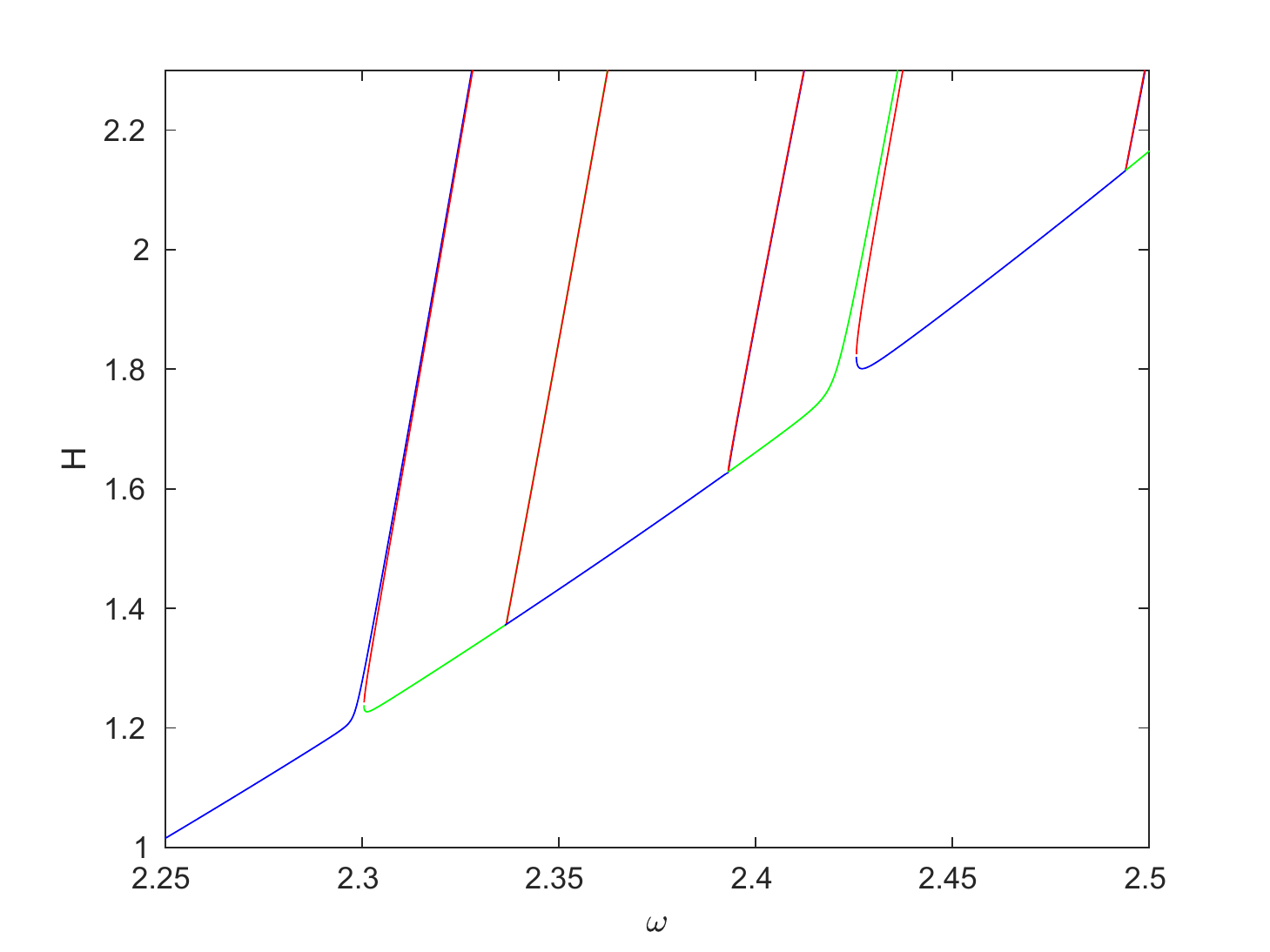}}\\
\subfloat[]
{\includegraphics[width=0.5\textwidth]{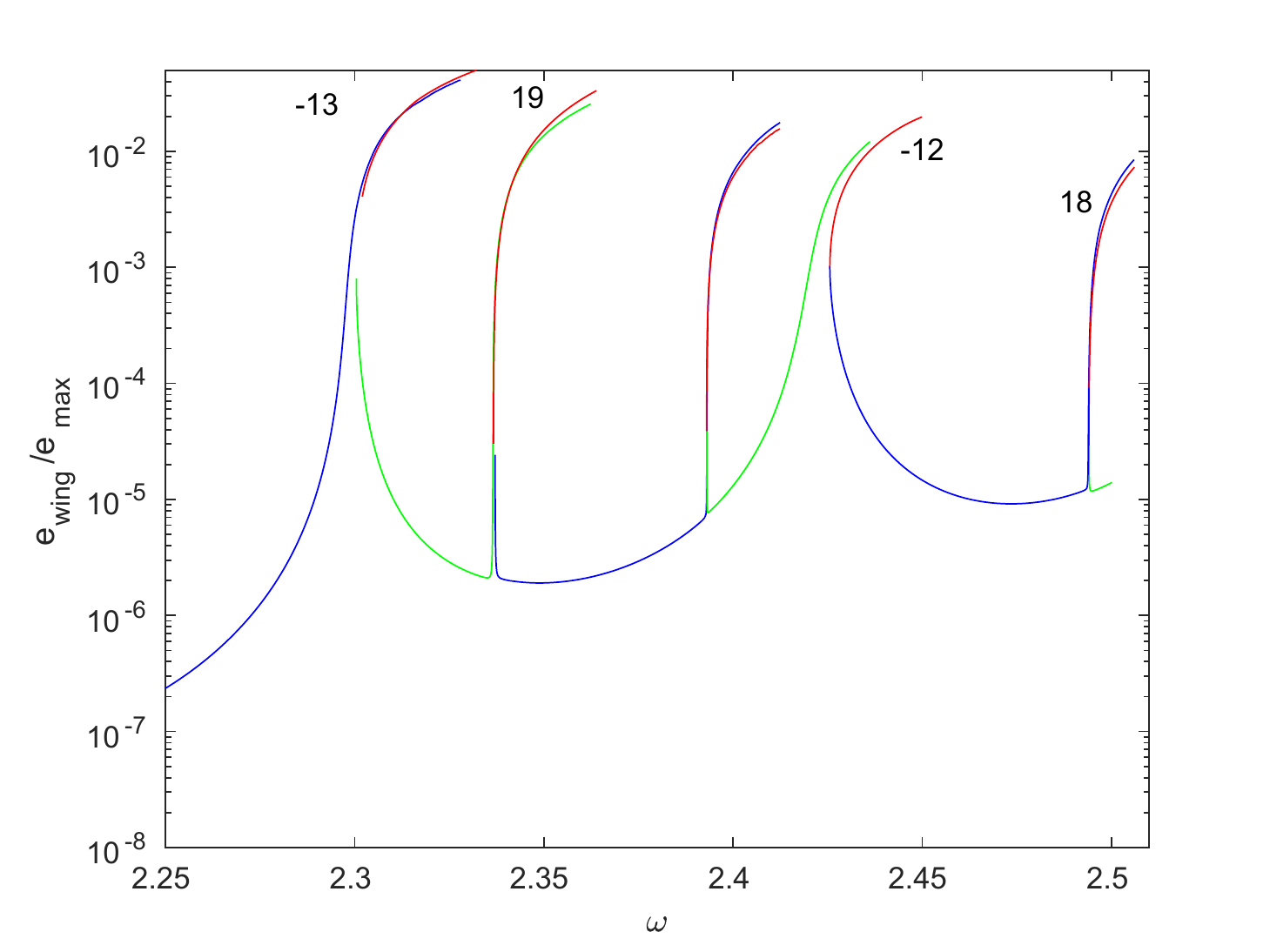}}
\subfloat[]
{\includegraphics[width=0.5\textwidth]{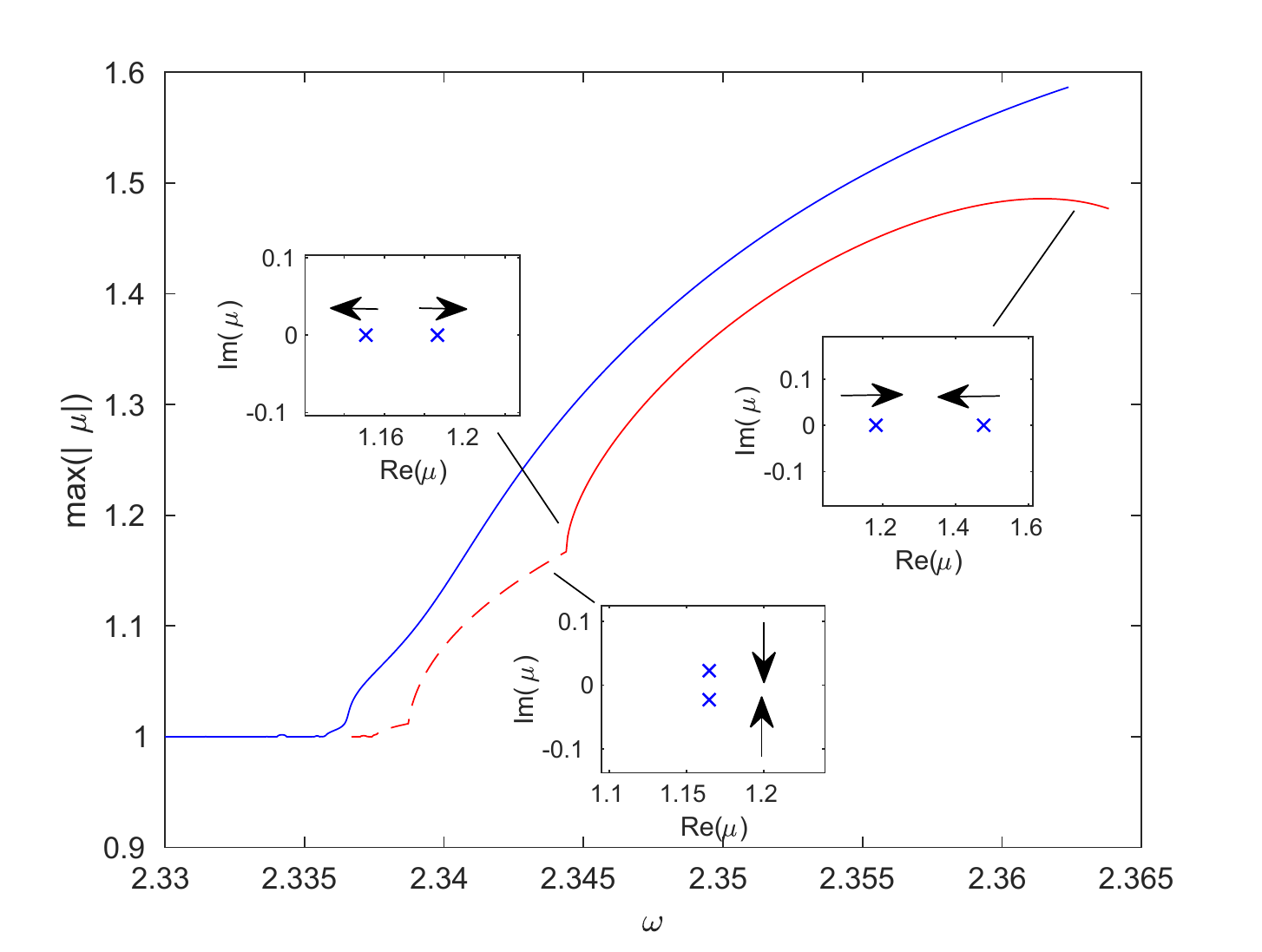}}
\caption{\footnotesize (a) Energy $H$, (b) the normalized average site energy $e_{wing}/e_{max}$ of the wings and (c) maximum moduli of the Floquet multipliers $\mu$ along different branches as functions of $\omega$ at $V_1 = 2/5$ and $N=60$ near the resonance $\omega = 2.337$. The Floquet multiplier with the maximum modulus has nonzero real and imaginary parts along dashed portions of the curve and is real along the solid one. Insets illustrate that transitions between these regimes occur due to the collisions of pairs of real and complex Floquet multipliers. Different colors correspond to different branches in (a). The numbers in (b) are the values of $m$ for each resonance (see the text for details). }
\label{fig:m60v2_5_full}
\end{figure}

Panel (c) of Fig.~\ref{fig:m60v1_3_full} shows the maximal moduli of the Floquet multipliers associated with the computed breathers near the resonance $\omega = 2.237$ as a representative example. As the top branch nears a resonance, a \emph{real instability}, which corresponds to a real Floquet multiplier $\mu>1$, manifests itself. As $\omega$ continues to increase along the branch, and the wings of the moving breathers become more pronounced, this real instability is accompanied by the emergence of \emph{complex instability} modes associated with Floquet multipliers $\mu$ that have nonzero imaginary part and satisfy $|\mu|>1$. As can be seen in the insets, the largest real multiplier is accompanied by a smaller real one that eventually collides with it. This collision results in the formation of a symmetric quadruplet of complex-valued multipliers. Meanwhile, both the bottom branch and the middle branch are stable near the resonance frequency.
Understanding the relevant turning point structure that connects the two is an interesting question for
future work.
In the case of the middle branch, stability only persists over a short interval of $\omega$, as complex instabilities quickly arise. In this case, the modulus of the complex instabilities is larger than that of any real instabilities that emerge. The real instabilities exist as pairs of real multipliers that collide, separate and rejoin, shifting between complex and real, similar to what is seen in the top branch. This behavior is demonstrated in panel (c) of Fig.~\ref{fig:m60v2_5_full}. The lower branch only becomes unstable as it merges with the top branch for the next resonance.

The results for $V=2/5$ and $N=60$ are shown in Fig.~\ref{fig:m60v2_5_full} and Table~\ref{table:FreqMult_v2_5}.
\begin{table}[!htb]
\centering
\begin{tabular}{ |M{2cm}|M{2.25cm}|M{1.25cm}|}
 \hline
  numer $\omega_m$ & approx $\omega_m$ & $m$ \\
 \hline
 $2.337$ & $2.348$ & $19$ \\
 \hline
 $2.494$ & $2.510$ & $18$ \\
 \hline
 $2.423$ & $2.434$ & $-12$ \\
 \hline
 $2.300$ & $2.306$ & $-13$ \\
 \hline
\end{tabular}
\caption{Comparison of approximate and numerical resonance values $\omega_m$ for $V_1 = 2/5$ and $N=60$. The approximate values are computed using \eqref{eq:Resonance}. }
\label{table:FreqMult_v2_5}
\end{table}
Overall, they are similar to the case $V_1 = 1/3$, but the number of resonances is smaller over the same interval of $\omega$. In both examples, one of the resonances is not accounted by \eqref{eq:Resonance}.
As can be seen in the respective figures, both positive and negative integer resonances manifest themselves
sequentially in the context of \eqref{eq:Resonance}, yet one cannot be included in this sequence. This
also constitutes an intriguing question for future study.
Additionally, in panel(c) of Fig.~\ref{fig:m60v2_5_full}, the maximum moduli of Floquet multipliers along the middle branch which satisfy $|\mu|>1$ are determined by two pairs of complex Floquet multipliers that start near the point $\mu=1$, in contrast to the case discussed above. These complex multipliers eventually collide to form a pair of real multipliers, which initially separate but then start moving toward each other, as shown in the insets.

\section{Dynamical consequences of real instabilities}
\label{sec:instab}
We now consider the consequences of the instability of a moving breather with real Floquet multipliers $\mu>1$. To this end, we perturb the breather along the corresponding eigenmode by solving Eq.~\eqref{eq:EoM} with the initial displacement vector set to $\mathbf{q}(0)+\epsilon \delta\mathbf{q}$ and initial momentum to $\mathbf{p}(0) + \epsilon \delta\mathbf{p}$, where $\mathbf{p}(t)$ and $\mathbf{q}(t)$ are the displacement and momentum vector functions, respectively, for the moving breather, $\delta\mathbf{q}$ and $\delta\mathbf{p}$ are the displacement and momentum parts of the unstable eigenmode, and $\epsilon$ measures the strength of the applied perturbation along
this unstable eigendirection.

\begin{figure}[!htb]
\centering
\subfloat[]{
\includegraphics[width=0.5\textwidth]{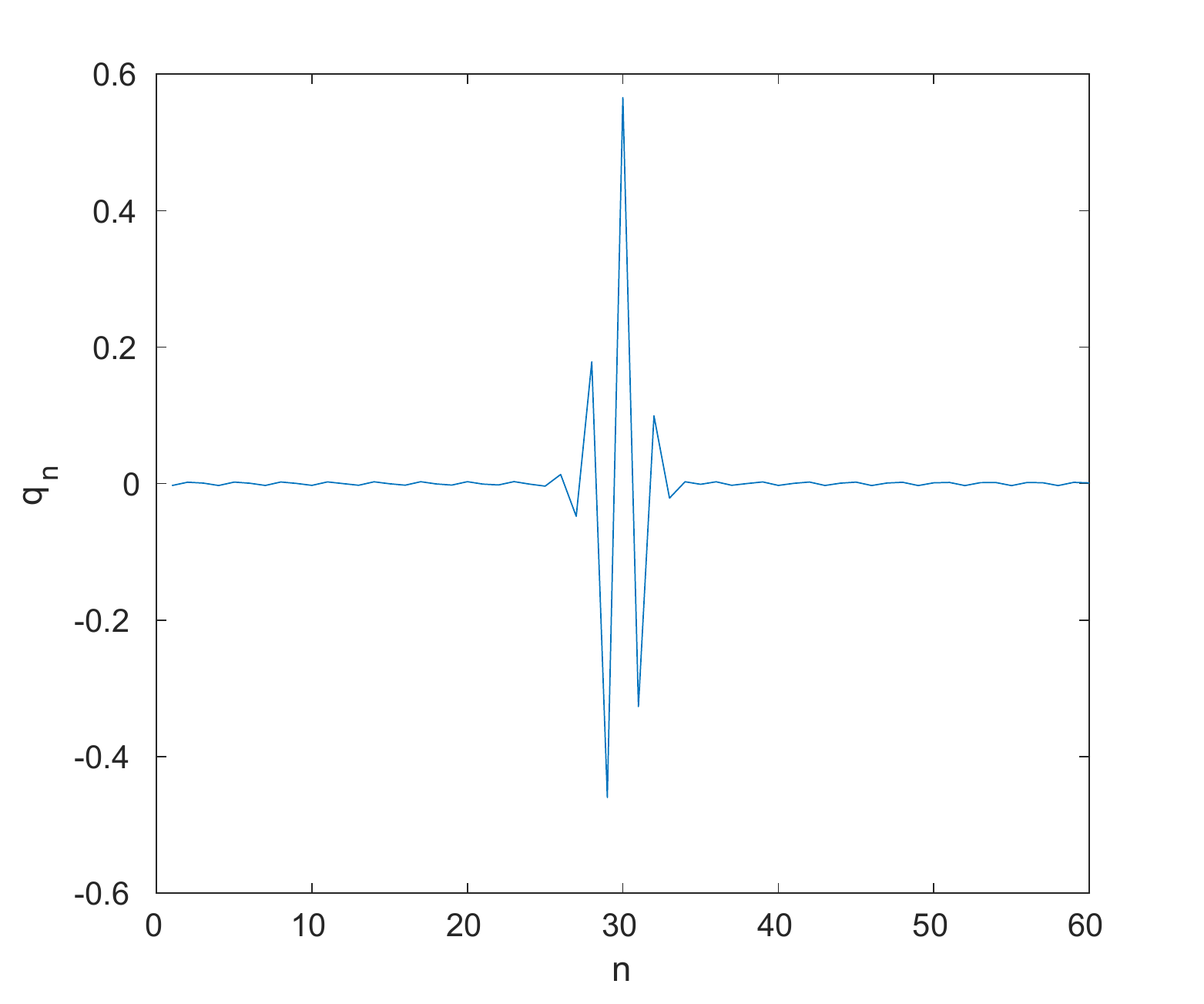}
}
\subfloat[]{
\includegraphics[width=0.5\textwidth]{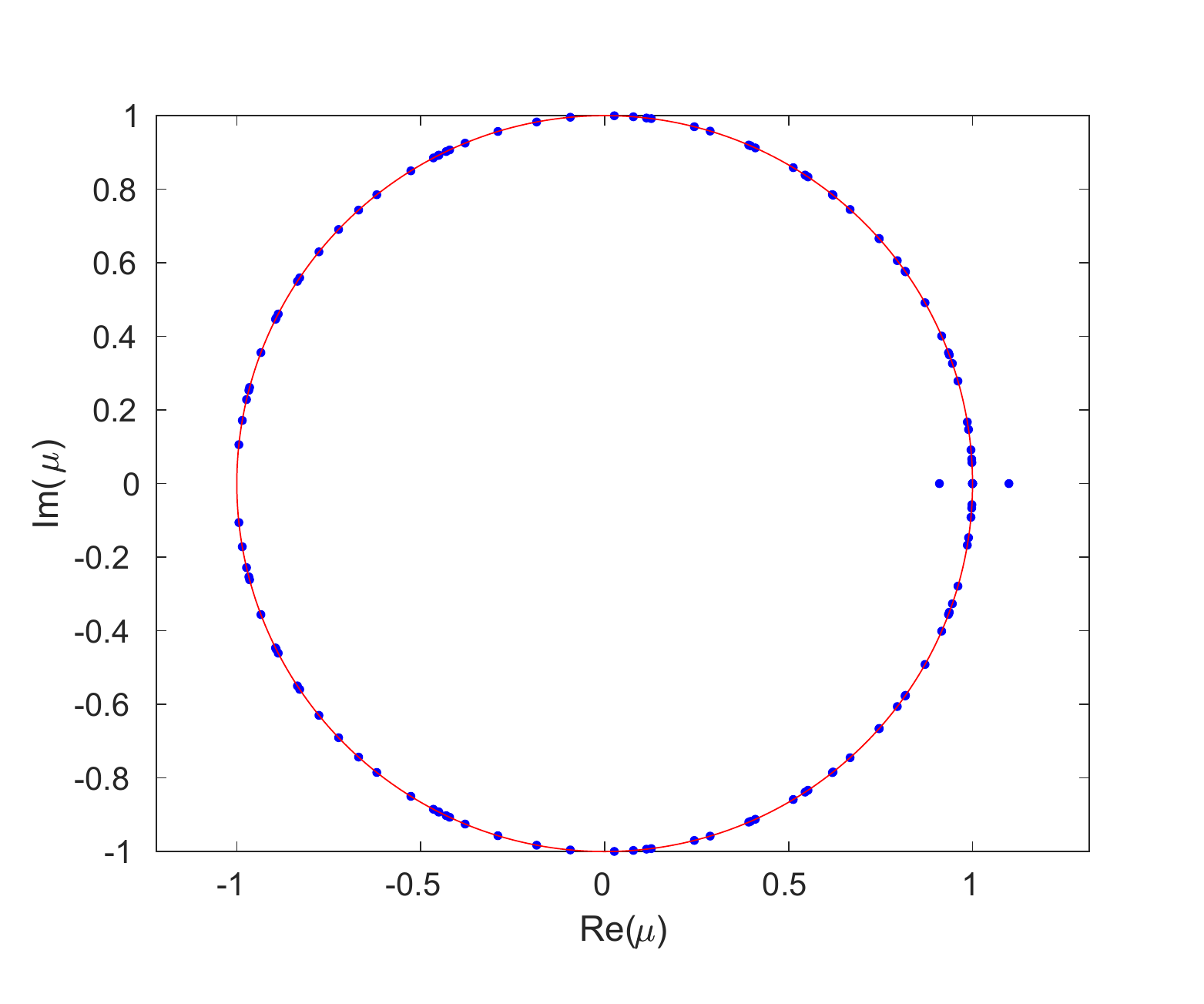}
}
\caption{\footnotesize (a) Displacement profiles $q_n$ of the unperturbed moving
  breather with $V_1 = 1/3$, $\omega = 2.424$, and $N=60$. (b) Floquet multipliers $\mu$.
  The largest real multiplier is $\mu = 1.0989$.}
\label{fig:m60_v1_3_unpert}
\end{figure}

\begin{figure}[!htb]
\centering
\subfloat[]{
\includegraphics[width=0.5\textwidth]{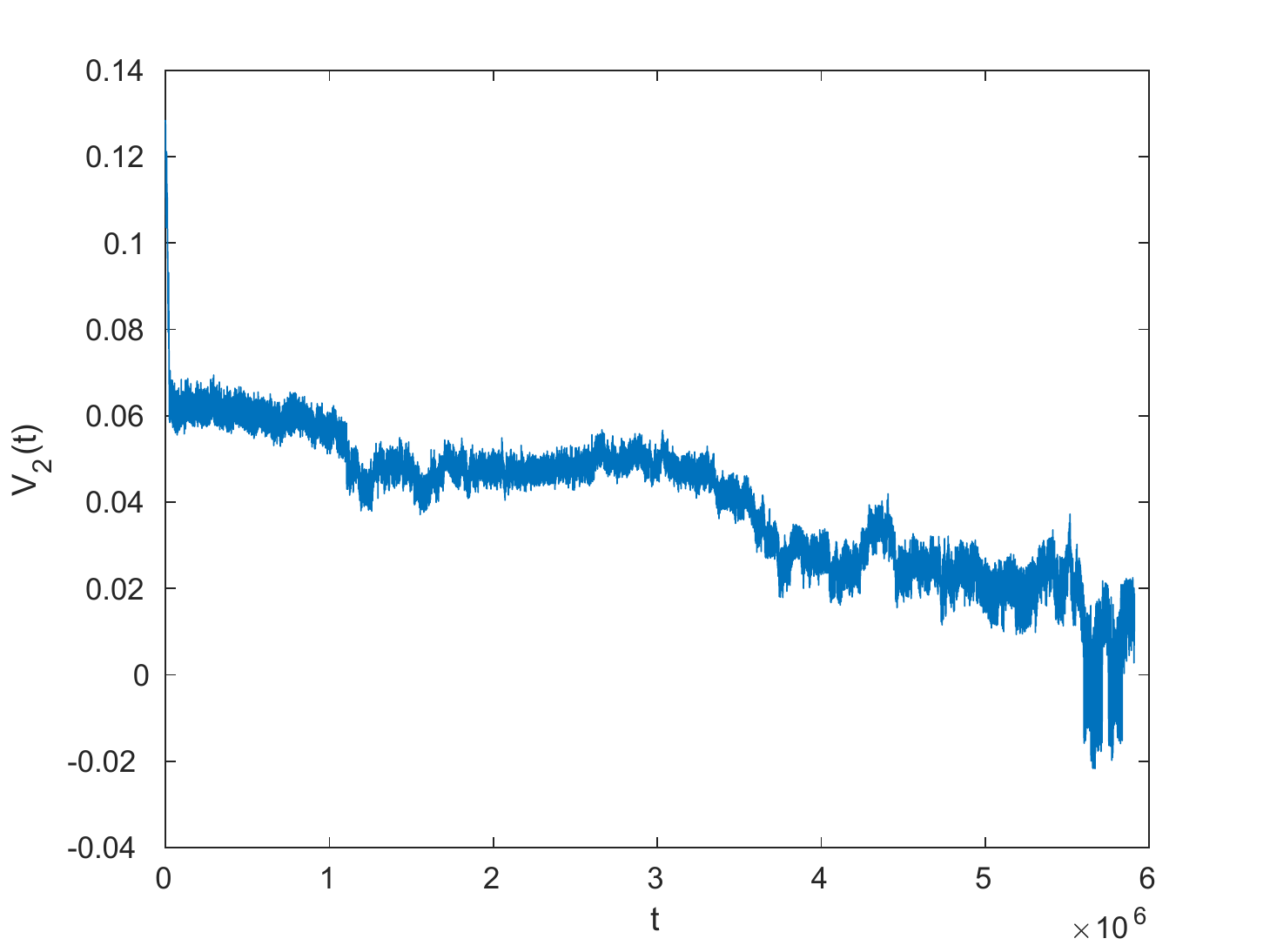}
}
\subfloat[]{
\includegraphics[width=0.5\textwidth]{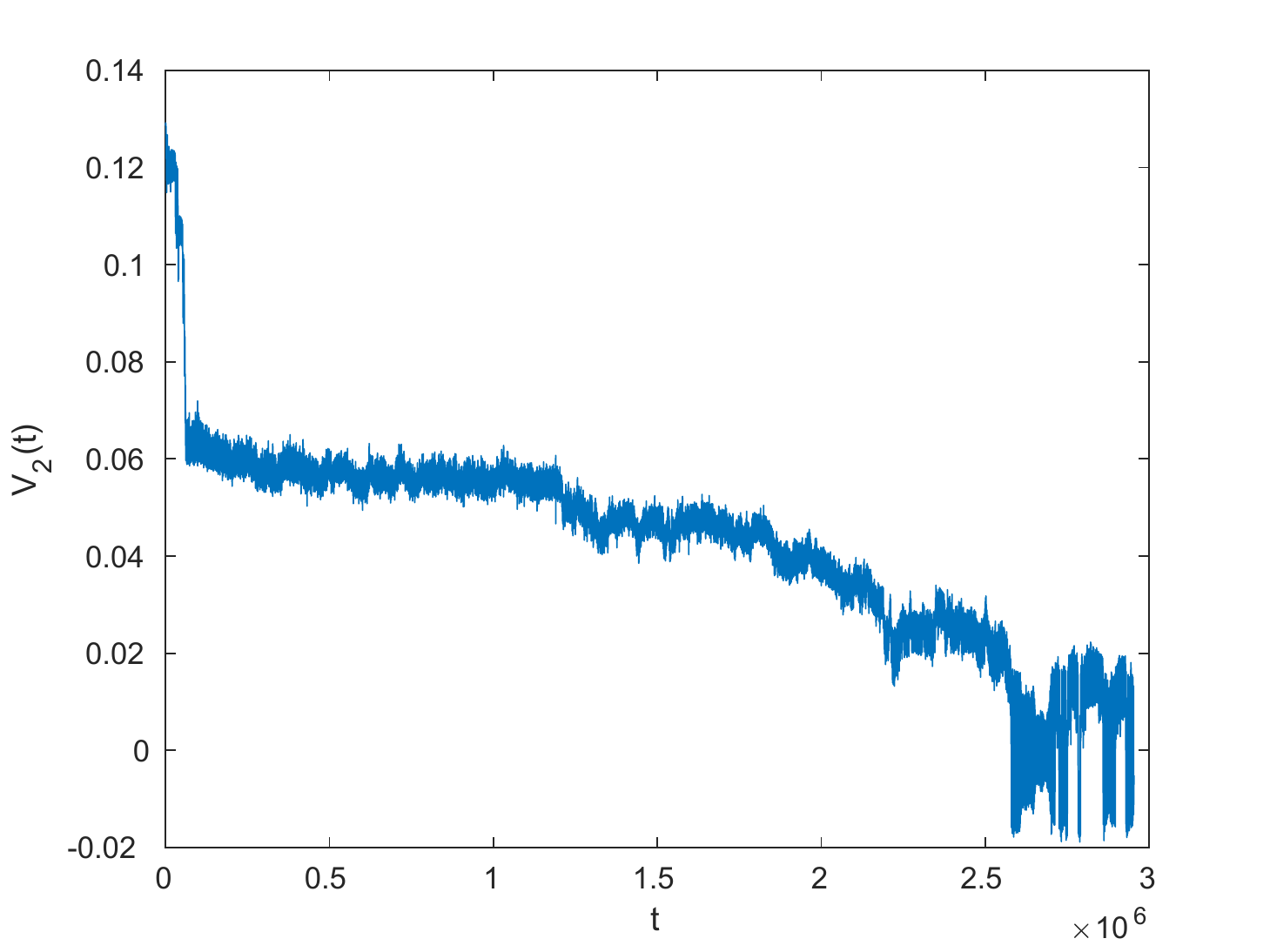}
}
\caption{\footnotesize Time evolution of the translational velocity $V_2$ for the moving breather with largest real Floquet multiplier $\mu=1.0989$ at (a) $\epsilon=-0.01$; (b) $\epsilon=0.01$. Here $V_1=1/3$, $\omega = 2.424$ and $N=60$. After an initial transient resulting from the instability manifestation,
  the breather can be seen to incur a very slow velocity decrease over the long time evolution.}
\label{fig:m60_v1_3_velocity}
\end{figure}

We consider the unstable moving breather with $V_1 = 1/3$, $\omega = 2.424$ and $N=60$, from the top (blue) branch in Fig.~\ref{fig:m60v1_3_full}, which has the maximum real Floquet multiplier $\mu=1.0989$ (see Fig.~\ref{fig:m60_v1_3_unpert}). Note that the breather has wings of relatively small amplitude. Fig.~\ref{fig:m60_v1_3_velocity} shows the evolution of the translational velocity $V_2$ when the breather is perturbed with $\epsilon = -0.01$ (panel (a)) and $\epsilon = 0.01$ (panel (b)). In both cases, after initial transients
leading to substantial deceleration, the velocity of the perturbed breather appears to stabilize and oscillate around specific values, before decreasing again and eventually coming to oscillate around zero.
As an inspection of the relative sizes of the horizontal and vertical axes reveals, this is a particularly
slow process.
Interestingly, the $\epsilon = -0.01$ perturbation case takes much longer to reach this state.

\begin{figure}[!htb]
\centering
\subfloat[]{
\includegraphics[width=0.5\textwidth]{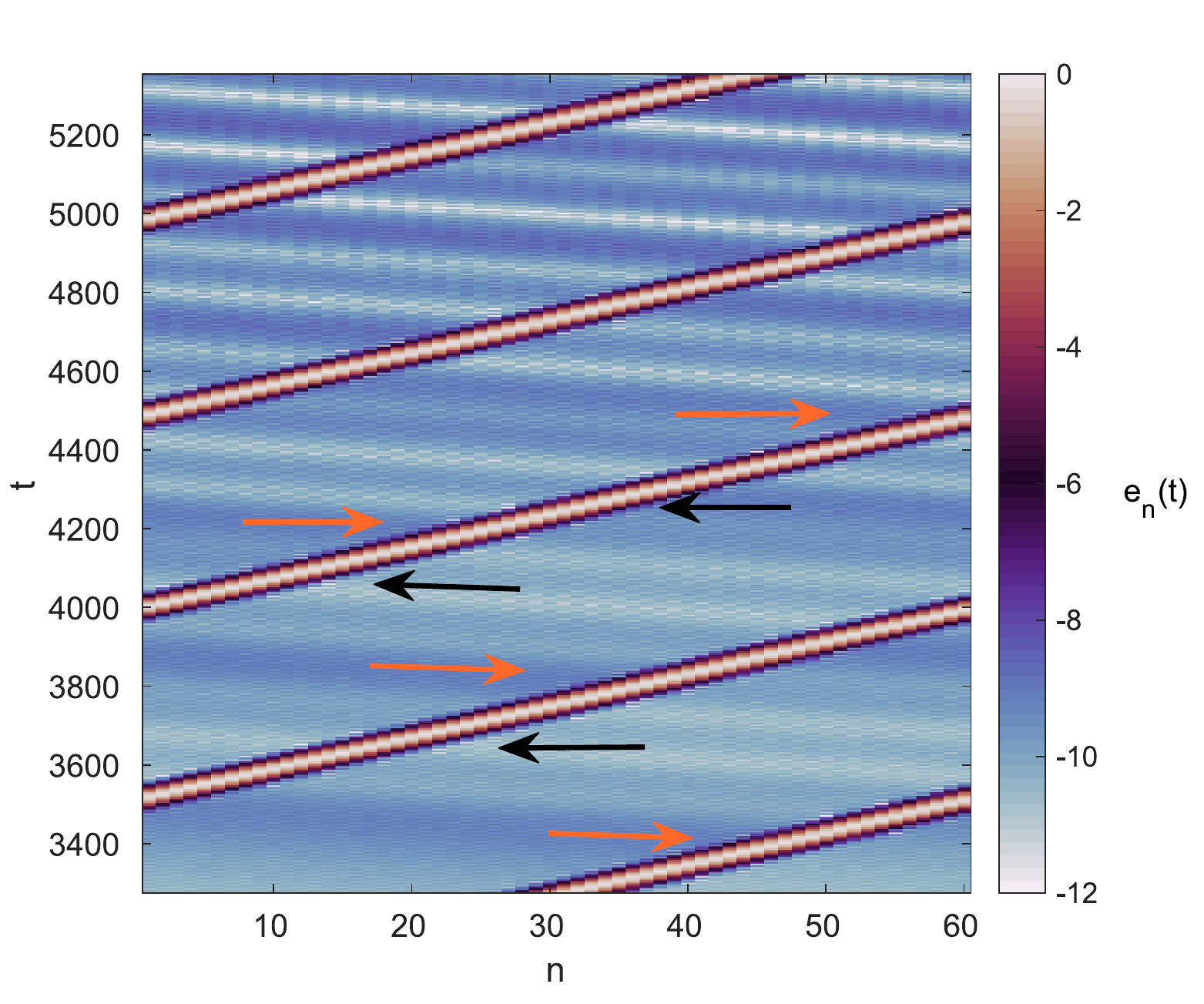}
}
\subfloat[]{
\includegraphics[width=0.5\textwidth]{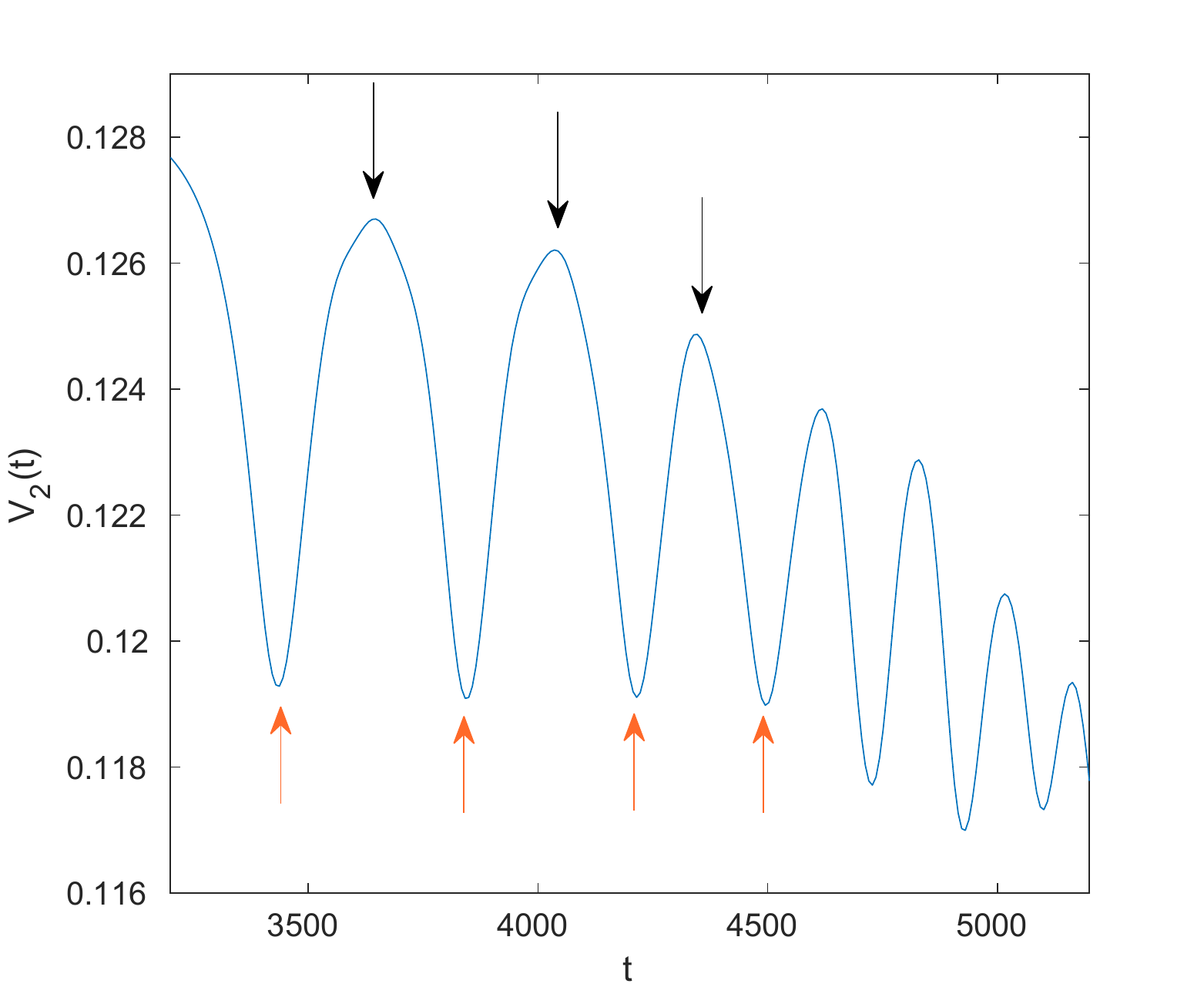}
}
\caption{\footnotesize (a) Space-time evolution of the site energy $e_n(t)$ and (b) time evolution of the velocity $V_2$ near the start of the simulation with $\epsilon=0.01$. The arrows pointing left and right in (a) correspond to the arrows pointing up and down, respectively, in (b). Here $V_1 = 1/3$, $\omega = 2.424$ and $N=60$.  }
\label{fig:m60_v1_3_initial}
\end{figure}

Figure~\ref{fig:m60_v1_3_initial} shows the space-time evolution of the energy density at the lattice nodes early on in the simulation for the case when $\epsilon = 0.01$. As can be seen in Fig.~\ref{fig:m60_v1_3_initial}(a), the core of the perturbed breather emits a backwards traveling wave. This corresponds to a minimum in the translational velocity $V_2$ as can be seen in Fig.~\ref{fig:m60_v1_3_initial}(b). Once this offshoot wave travels around the chain of particles and strikes the core, a secondary wave is emitted. This additional wave travels around the chain and its collision with the core is associated with a maximum in $V_2$ as can be seen in Fig.~\ref{fig:m60_v1_3_initial}(b). As more and more waves are emitted, the time between successive extrema decreases due to more frequent collisions. Consequently, the oscillation of $V_2$ becomes more and more pronounced. Nevertheless, this phenomenology reflects the instability manifestation
and explains the progressive decrease of the energy harbored within the breather the corresponding increase of
energy redistributed throughout the lattice.
%


\section{Conclusions}
\label{sec:conclusions}

In the present work we have revisited the topic of the identification, stability classification and dynamical instability
manifestation of discrete breathers in the well-known $\beta$-FPU lattice. Our exploration has enabled a number
of insights into this problem. In particular, we developed a numerical procedure of continuation along a sequence of rational values of the
period-wise velocity that allows for the examination of different breather families traversing $r$ sites of the lattice over
$s$ multiples of the breather period. The continuation of the relevant waves over the frequency of the breather
revealed an intriguing resonance structure, as well as the multivalued nature of the corresponding energy-vs-frequency
dependence, enabling the identification of multiple breather waveforms for the same frequency. The resonance
structure  was elucidated quantitatively for different integer harmonics of frequencies around the breather
in comparison with ones of the continuous spectrum, following the work of~\cite{YoshiDoi07}. The specific
harmonics leading to the observed resonances were explicitly identified. At the stability level, the Floquet
multipliers of the different branches involved in the resonances were discussed, including also their
potential collisions and bifurcations in the complex plane. We remark that in contrast to stationary breathers \cite{jcmprl}, the emergence of real instability in this case was not associated with the change in the monotonicity of energy as a function of frequency.
Finally, long-time simulations of the dynamical evolution were performed using a symplectic method in order to reveal the manifestation of the relevant instabilities
(via the emission of and collision with offshoot waves) and their net result in decelerating and eventually
stopping the initially moving breather state.

Naturally, while this study has provided new insights into the dynamics of moving breathers, it has also raised some questions
that require further consideration. For instance, among the interesting technical questions
that arose were the specific bifurcation structure of the associated periodic orbits in the vicinity of
the highly nonlinear resonances that we explored. Another related aspect concerned the fact that
we could enumerate all positive and negative resonances in sequence, in connection with the
analytical condition of \eqref{eq:Resonance} but for a single one. It is also interesting to investigate whether the results obtained
in this work extend to Klein-Gordon lattices.

In addition, there exist larger scale questions for future studies. For instance, it would be interesting
to explore how the resonance structure and nonlinear state continuation would manifest themselves
in higher-dimensional models. In the latter, the issue of transverse (modulational along a stripe
or a ring) stability of the relevant waveforms would need to be considered as well. Another aspect of
consideration that at the moment eludes a systematic mathematical formulation is the existence
of traveling waveforms with genuinely real (rather than rational) period-wise velocity. Such questions are of substantial
interest for potential future investigations.


\appendix
\section*{Appendix: Additional traveling breather solutions}
\label{sec:doublebranch}


Investigations of the resonances suggest that a second set of solutions coexists along with the solutions discussed in the main text. These additional solutions can be found by employing a method similar to the one described in Sec.~\ref{sec:methods} for obtaining moving breathers from stationary breathers. By scaling the momentum vector of the moving breather solution and using this scaled momentum along with the unscaled displacement vector as an initial guess, Newton's method can be employed to obtain these secondary solution branches. The primary and secondary solutions typically differ in how the Floquet multipliers at the origin evolve after a resonance. In what follows, these dual solution sets are examined for the middle branch near the resonance at $\omega = 2.352$ when $N = 60$ and $V_1 = 1/3$.

\begin{figure}[!htb]
\centering
\subfloat[]
{\includegraphics[width=0.33\textwidth]{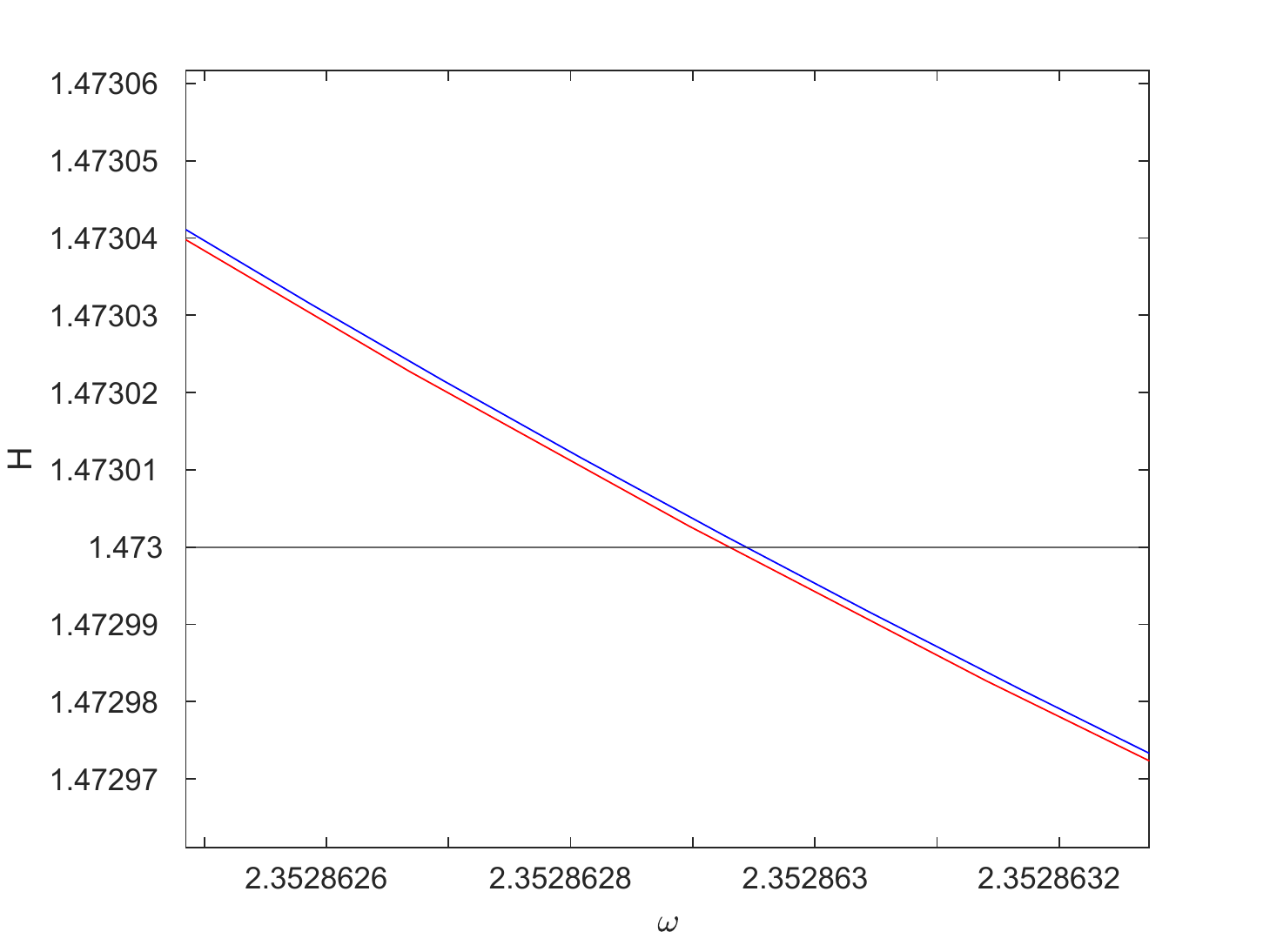}}
\subfloat[]
{\includegraphics[width=0.33\textwidth]{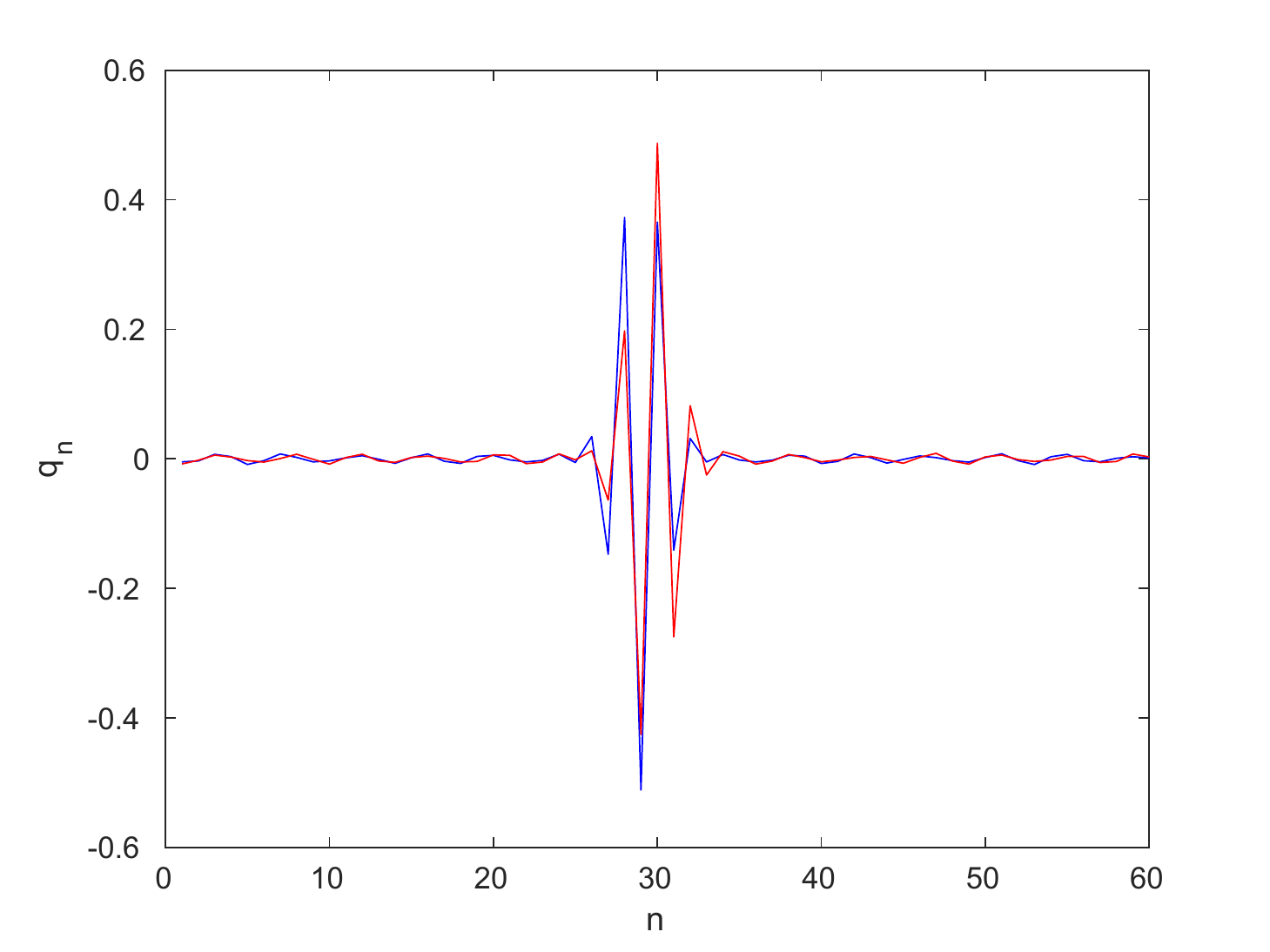}}
\subfloat[]
{\includegraphics[width=0.33\textwidth]{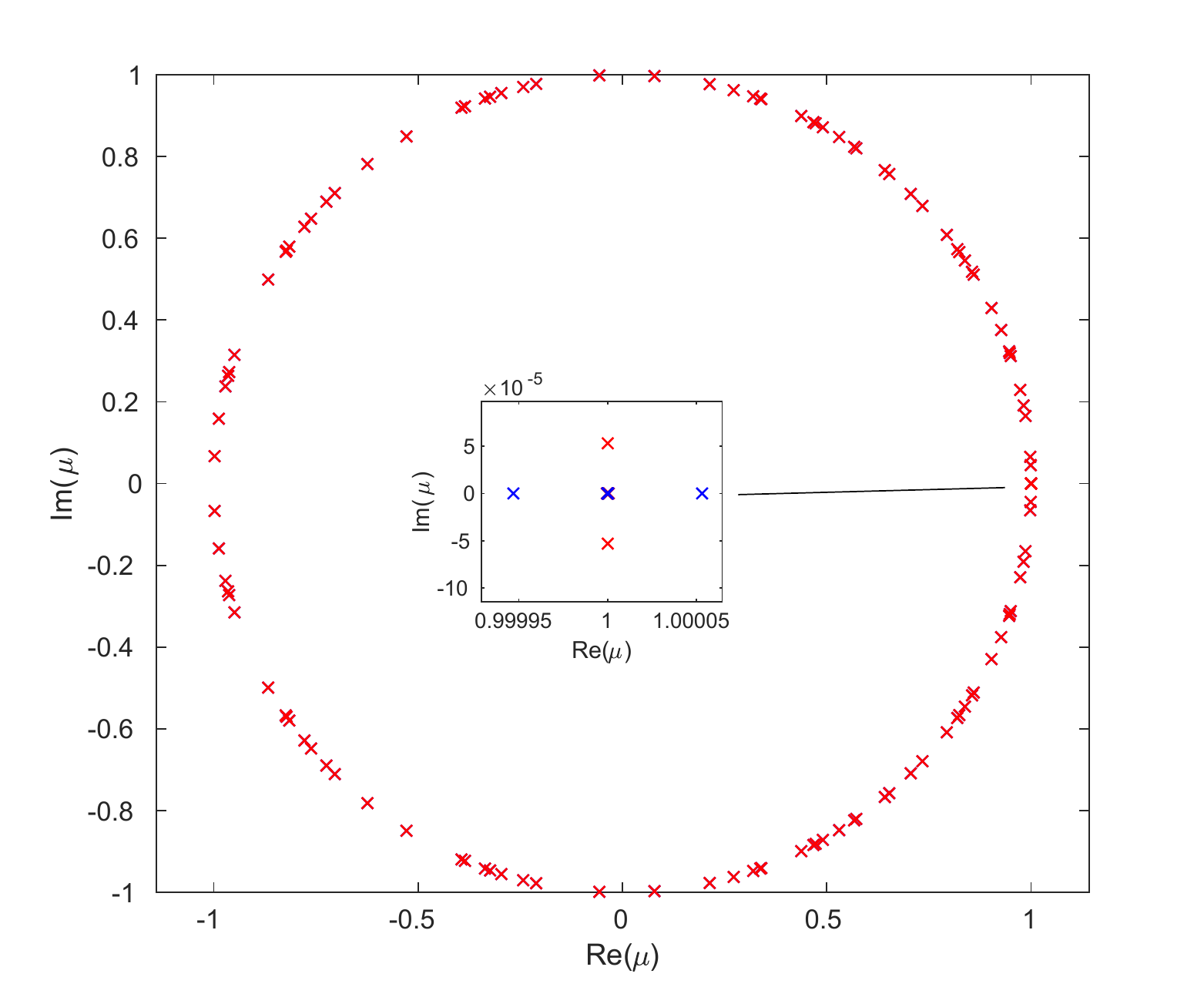}}\\
\subfloat[]
{\includegraphics[width=0.33\textwidth]{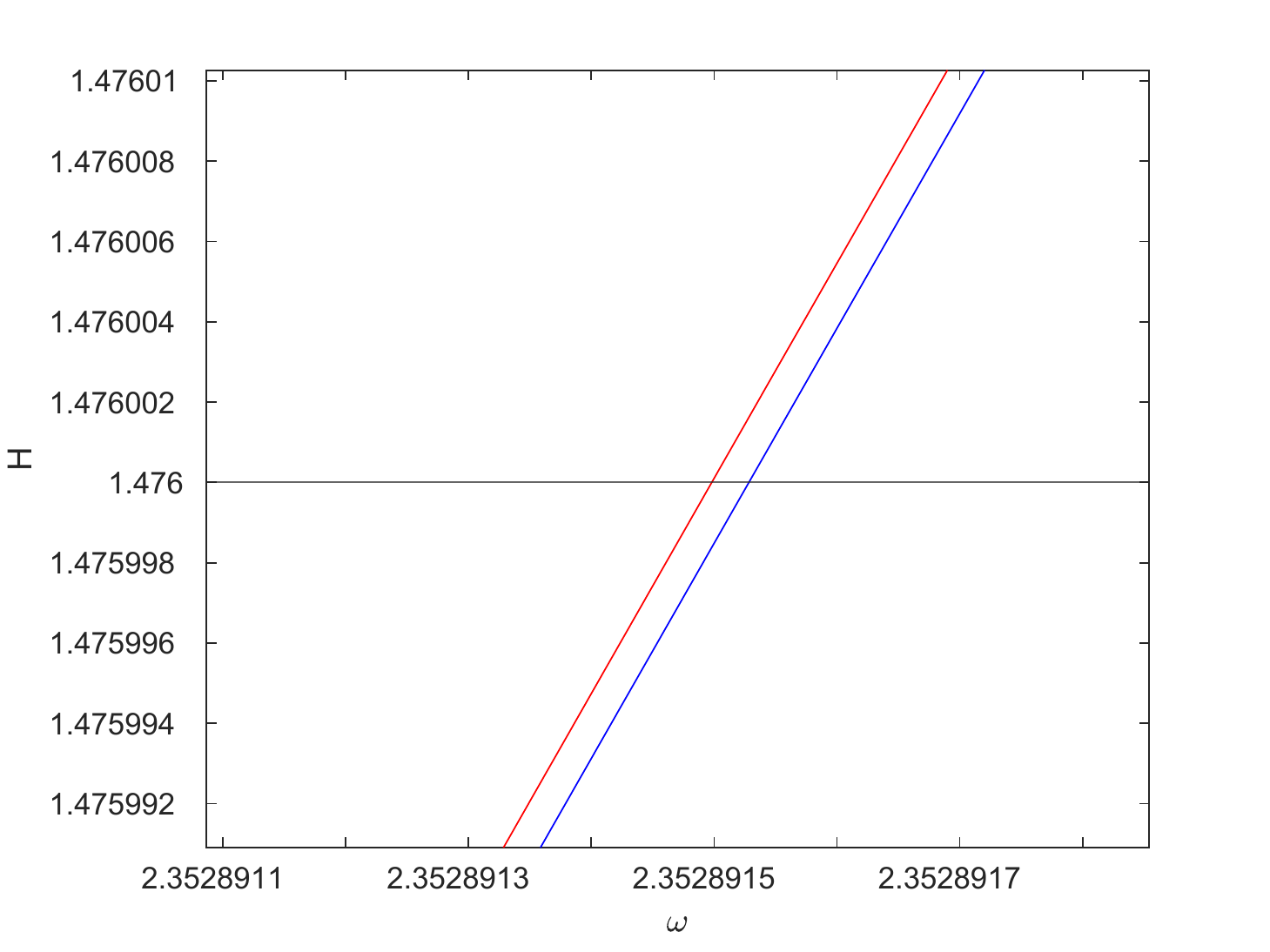}}
\subfloat[]
{\includegraphics[width=0.33\textwidth]{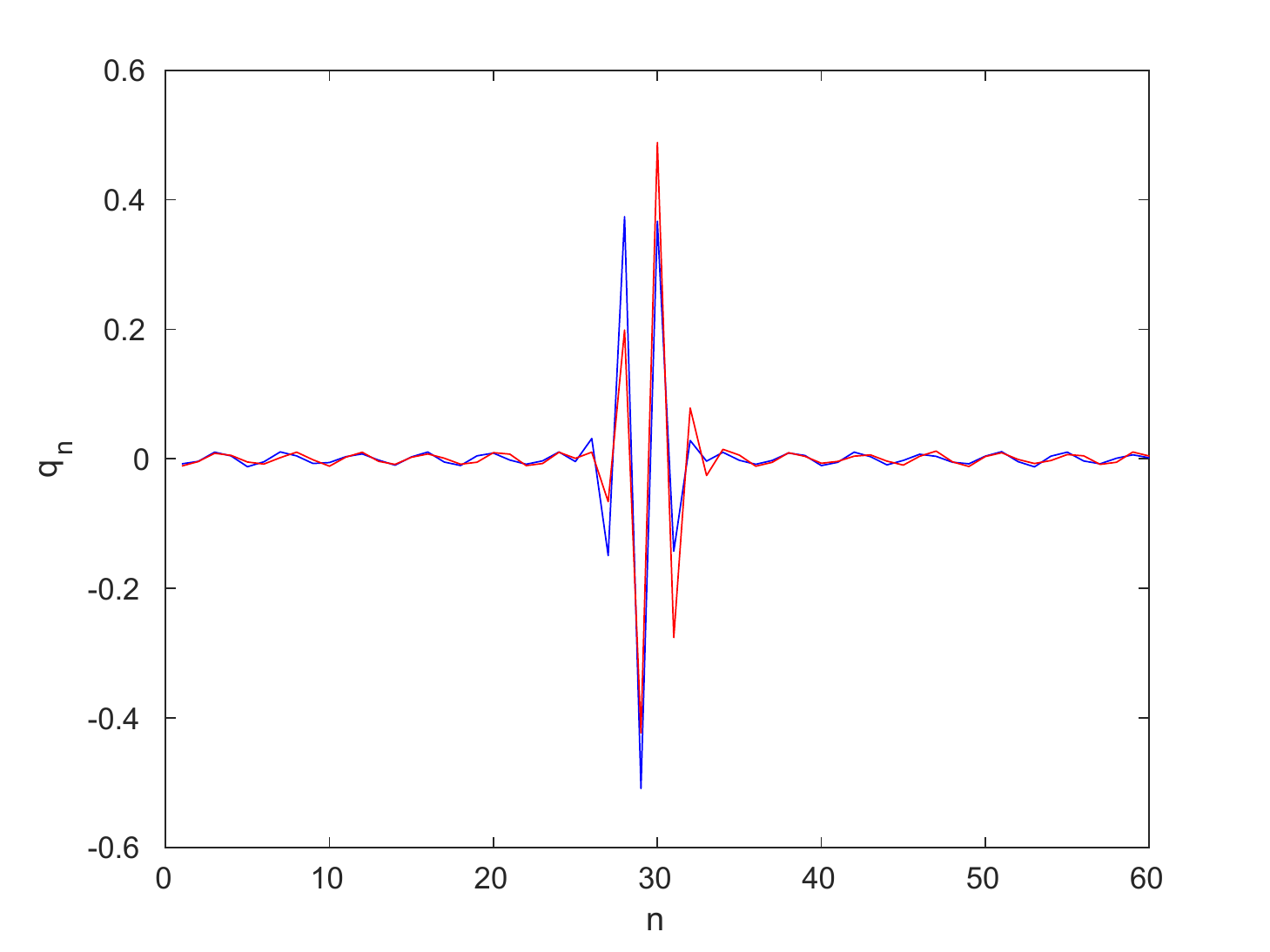}}
\subfloat[]
{\includegraphics[width=0.33\textwidth]{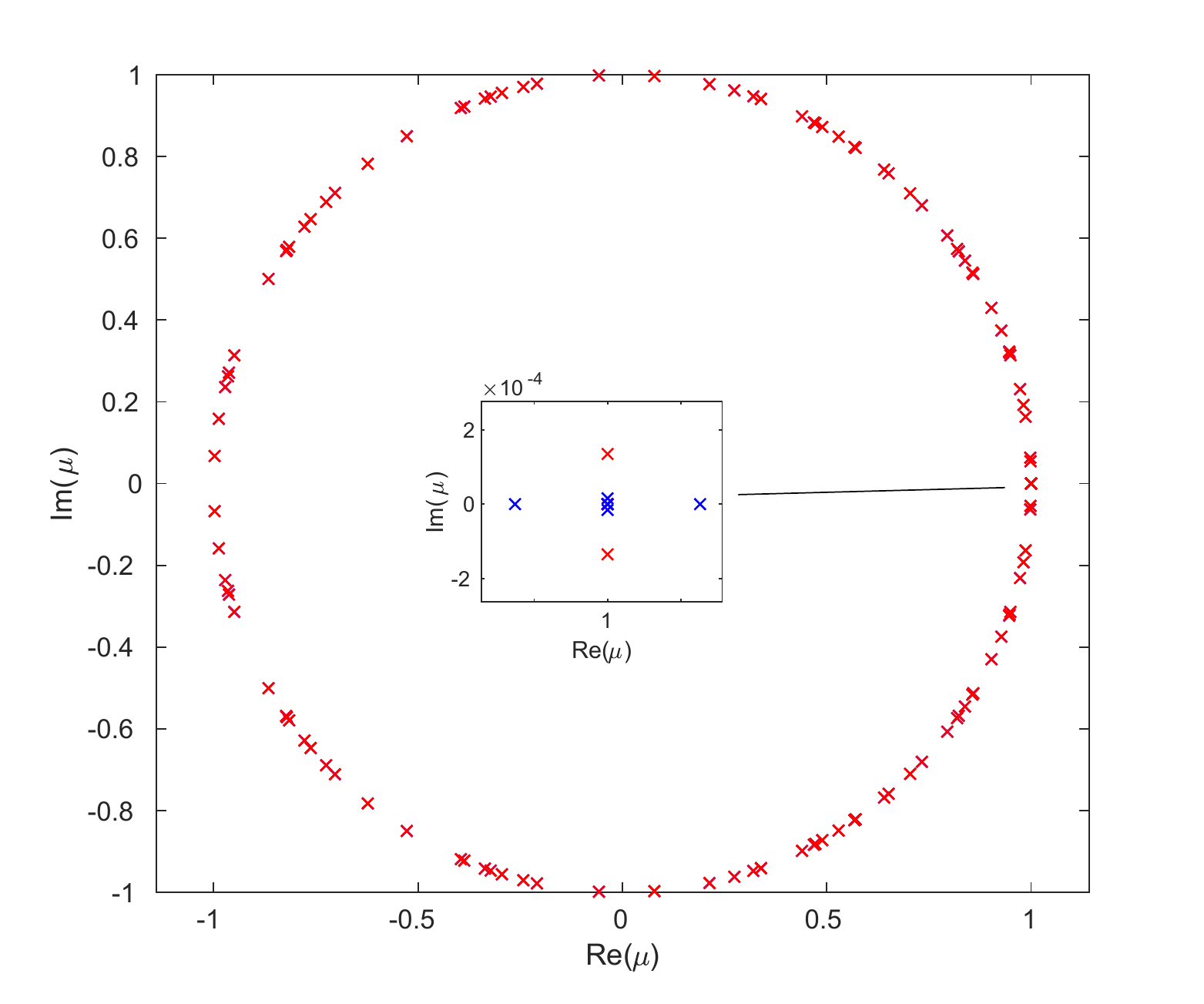}}\\
\subfloat[]
{\includegraphics[width=0.33\textwidth]{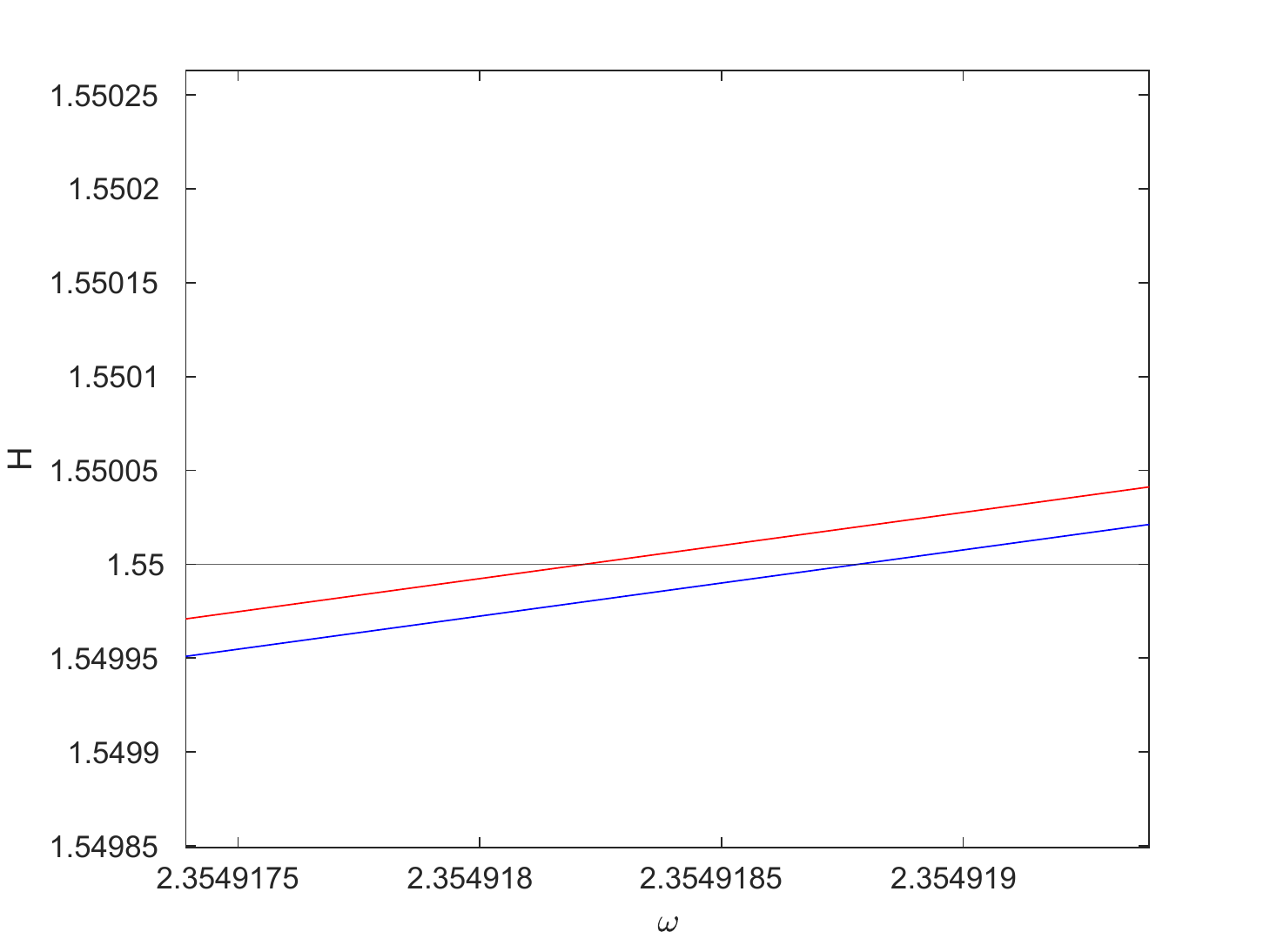}}
\subfloat[]
{\includegraphics[width=0.33\textwidth]{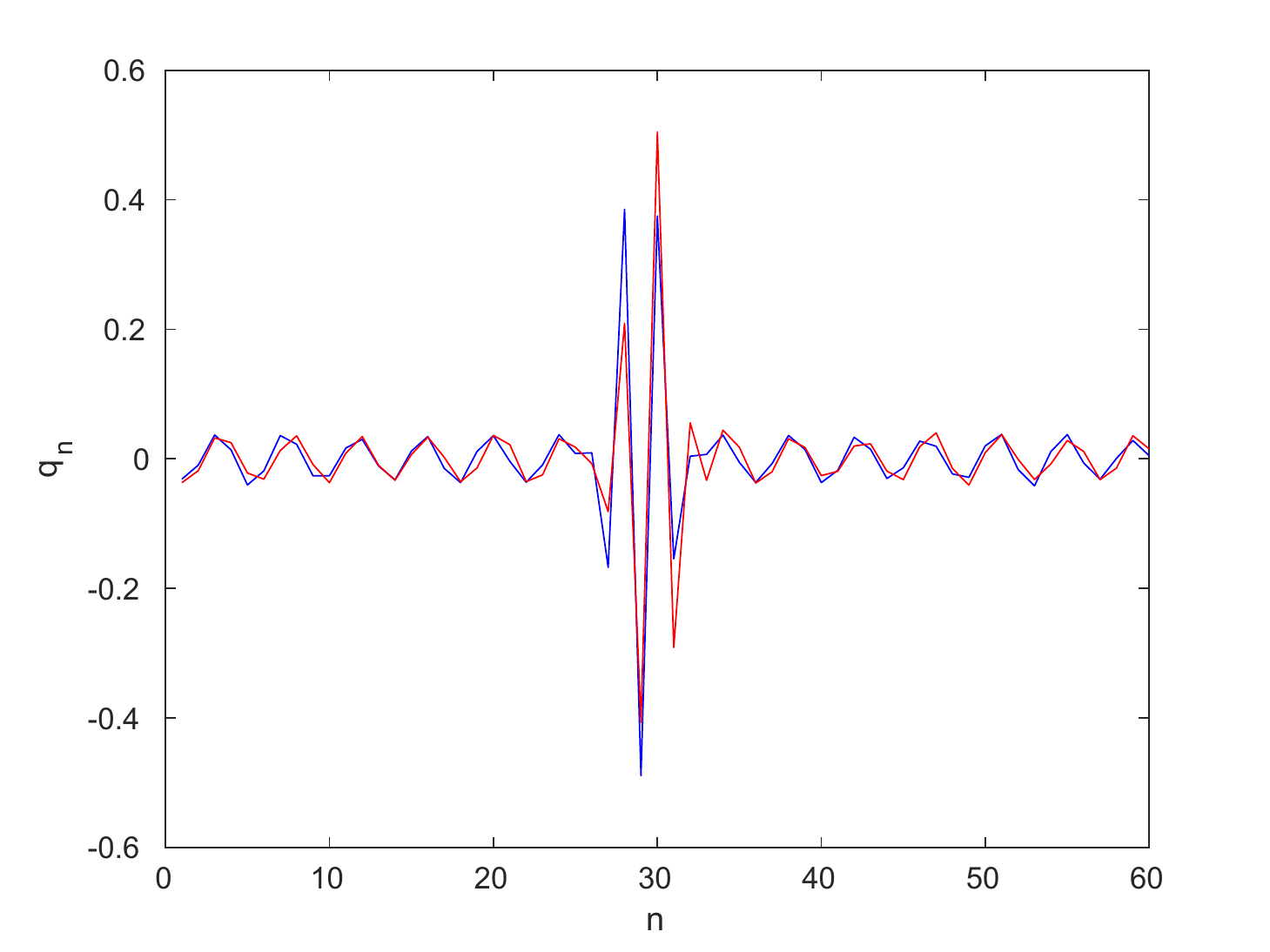}}
\subfloat[]
{\includegraphics[width=0.33\textwidth]{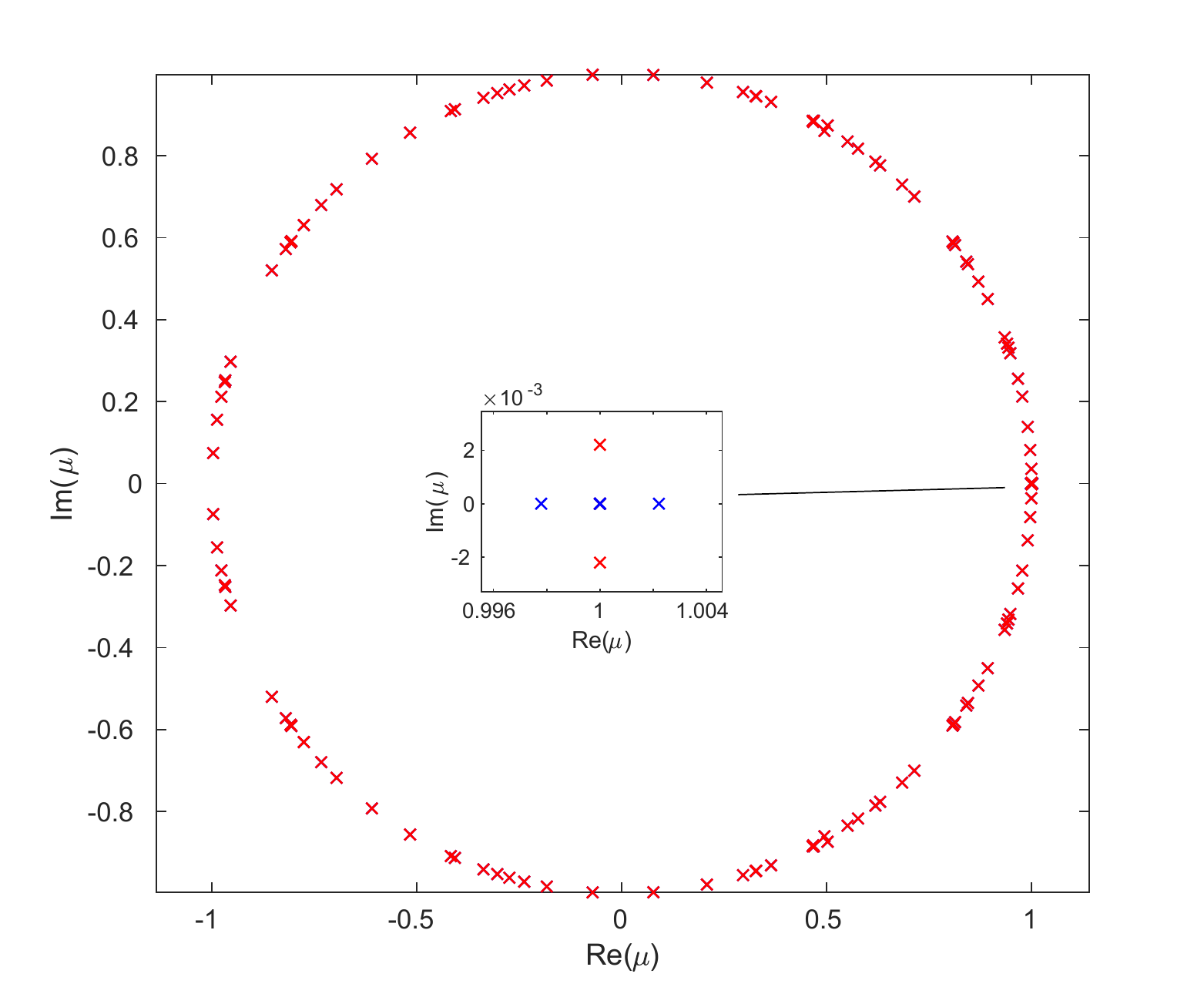}}
\caption{\footnotesize Coexisting solutions. Panels (a), (d) and (g) show the energy $H$ versus frequency $\omega$ along two coexisting solution branches, red and blue, near the resonance at $\omega = 2.352$. The horizontal black line marks the energy of the two solutions whose displacements $q_n$ are compared in panels (b), (e) and (h), respectively. Panels (c), (f) and (i) show the corresponding Floquet multipliers, with insets zooming in on the multipliers near $\mu=1$. In each case, a pair of Floquet multipliers is separating from the initial sextuple at $\mu=1$. Here $V_1 = 1/3$ and $N=60$.}
\label{fig:Res2355_Right}
\end{figure}
An example of this systematic comparison can be seen in Fig.~\ref{fig:Res2355_Right}. Panels (a), (d) and (g) of Fig.~\ref{fig:Res2355_Right} show the energy-frequency dependence along the two different solution branches (blue and red) near the resonance frequency. As the amplitude of the wings increases, the gap between the two solutions increases as well. Panels (b), (e) and (h) compare the displacements of the two solutions with the same energy (shown by horizontal line in panels (a), (d) and (g), respectively) and slightly different frequency. Note that the wings appear to be
essentially in phase with each other. We emphasize that the two solutions are not simply different time snapshots of the same breather. This can be seen by observing the difference in the Floquet multipliers near unity. The multipliers are depicted in panels (c), (f) and (i) for the pairs of solutions shown in panels panels (b), (e) and (h), respectively. Note that for the blue branch we see the emergence of two real Floquet mulitpliers that separate from the ones at $\mu=1$. However, for the red branch, the Floquet multipliers that leave $\mu=1$ move along the unit circle (rather than the
real axis). Thus, one solution branch develops a real instability associated with a small real multiplier, while the other does not. This is reminiscent of the commensurability effect discussed in Chapter 4 of \cite{Cretegny98} (see, for example, Fig. 4.11 therein).

\begin{figure}[!htb]
\centering
\subfloat[]
{\includegraphics[width=0.5\textwidth]{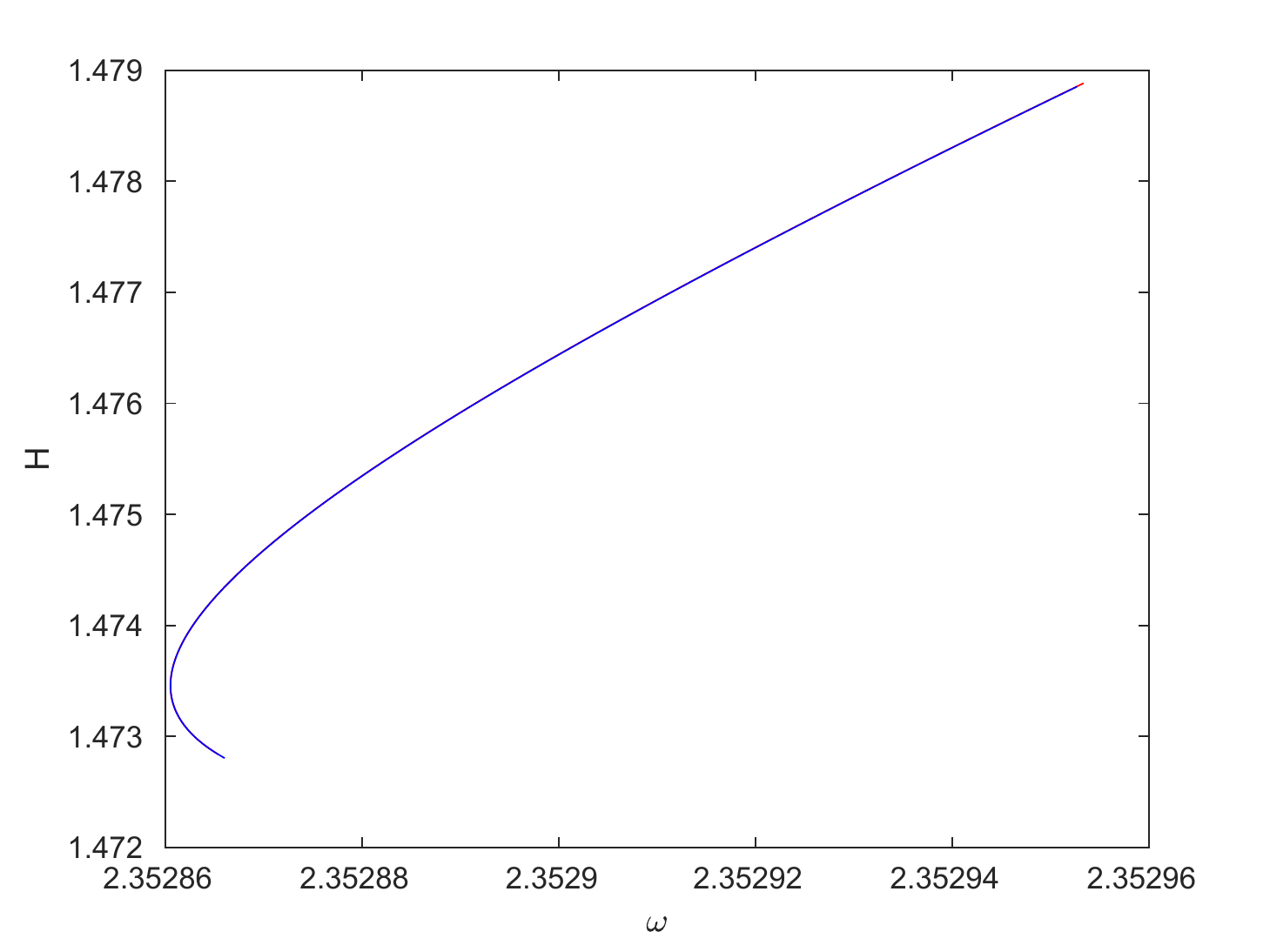}}
\subfloat[]
{\includegraphics[width=0.5\textwidth]{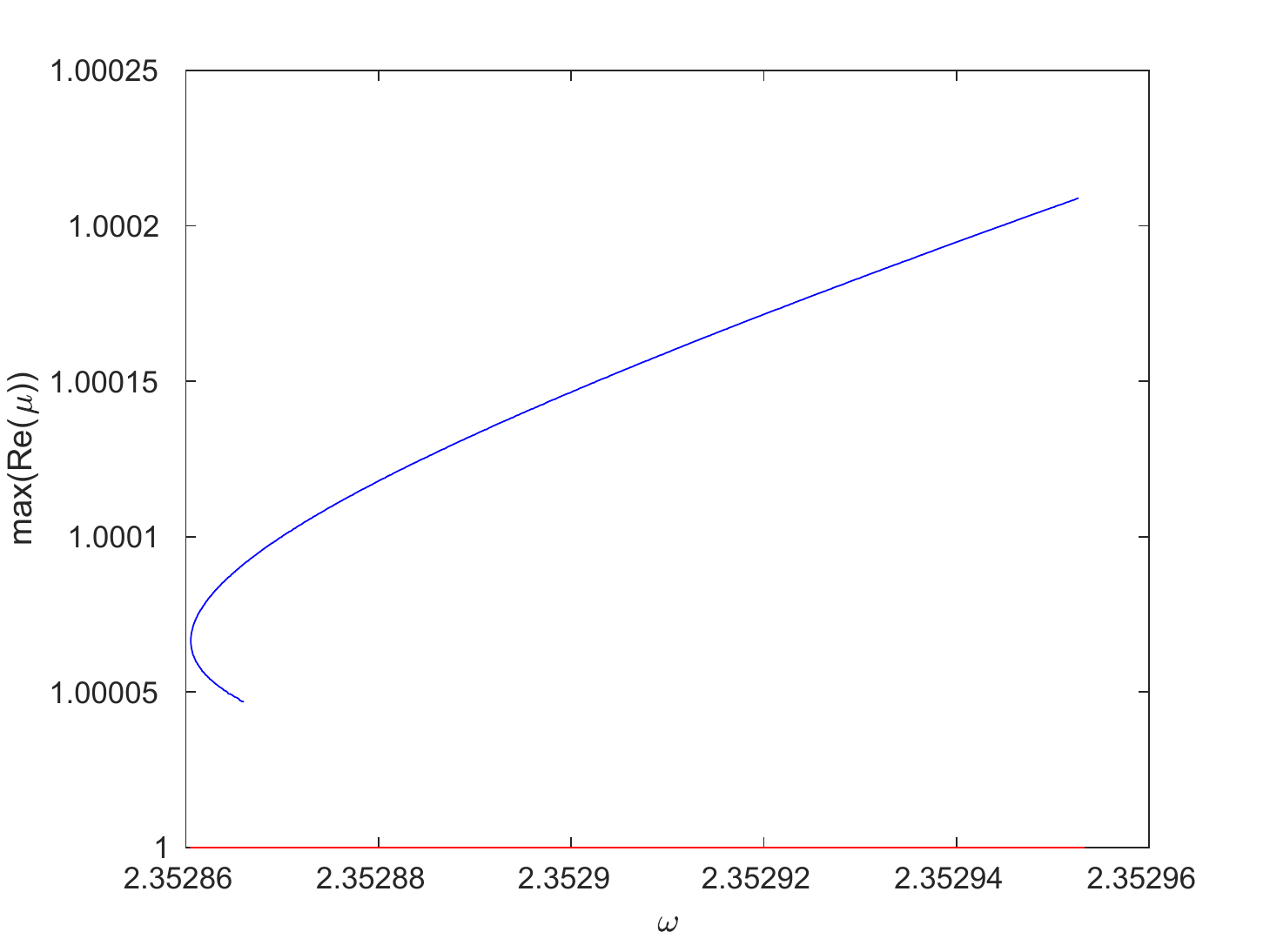}}
\caption{\footnotesize (a) Energy $H$ versus frequency $\omega$ and (b) the maximum real Floquet multipliers $\mu$ along the two solutions near the resonance at $\omega = 2.352$. The colors correspond to those used in Fig.~\ref{fig:Res2355_Right}. Here $V_1 = 1/3$ and $N=60$. In the left panel the blue and red branches cannot be distinguished over the scale of the figure (see
  also the magnified pictures in the left panels of Fig.~\ref{fig:Res2355_Right}). }
\label{fig:Res2355_TP}
\end{figure}
In Fig.~\ref{fig:Res2355_TP}, the dependence of the energy $H$ and maximum real Floquet multipliers $\mu$
on the breather frequency is shown near the turning point connecting the middle and bottom branches. The colors in each figure correspond to those used in Fig.~\ref{fig:Res2355_Right}. As can be seen, while the real Floquet multipliers along the red branch, in which the Floquet multipliers emerge along the unit circle, staying close to $\mu=1$, the largest real Floquet multiplier along the blue branch increases steadily as the energy increases. It should be explicitly
mentioned here that the energy of the two branches {\it cannot} be distinguished over the scale of the left panel.

%
\begin{figure}[!htb]
\centering
\subfloat[]
{\includegraphics[width=0.33\textwidth]{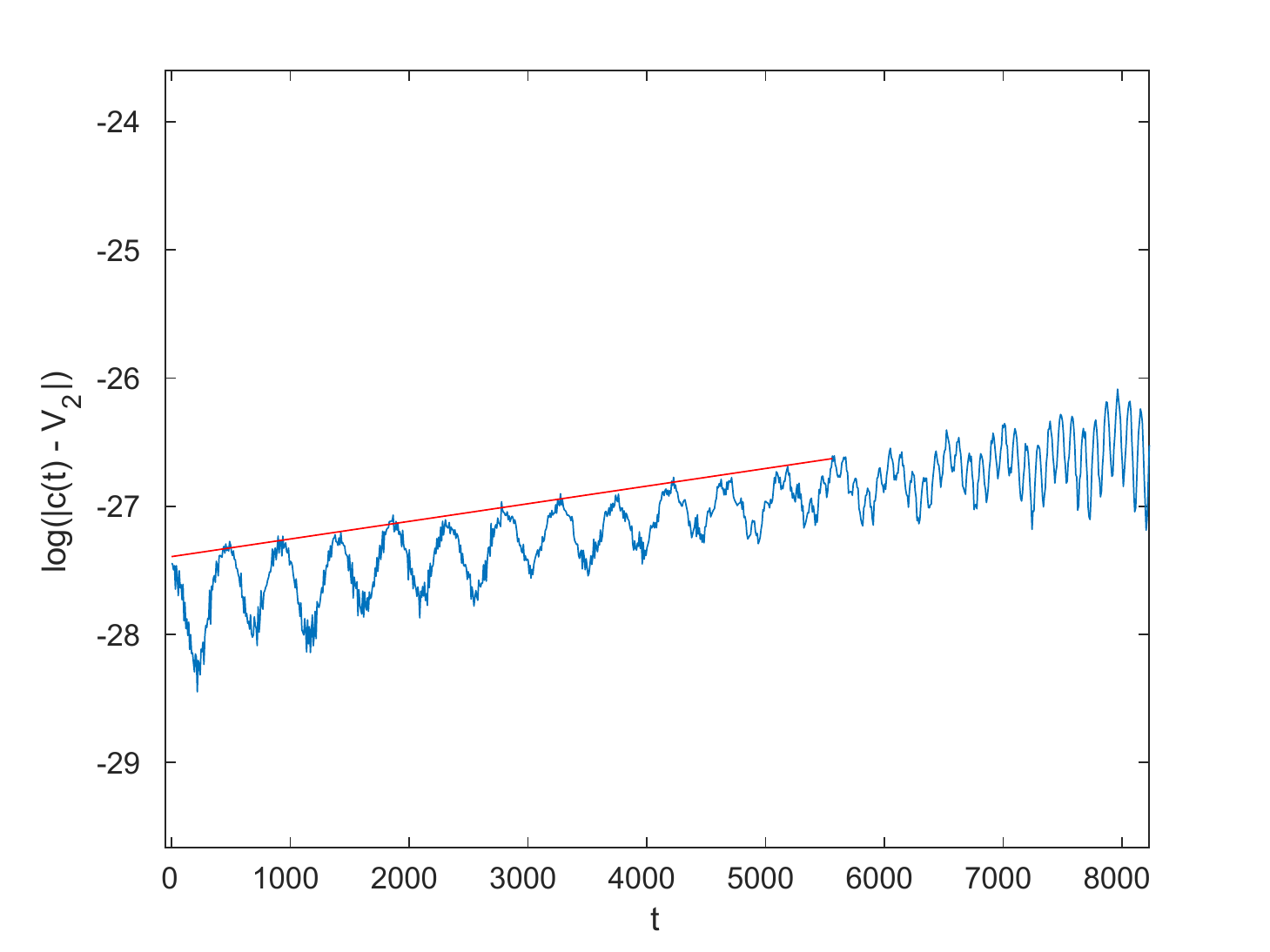}}
\subfloat[]
{\includegraphics[width=0.33\textwidth]{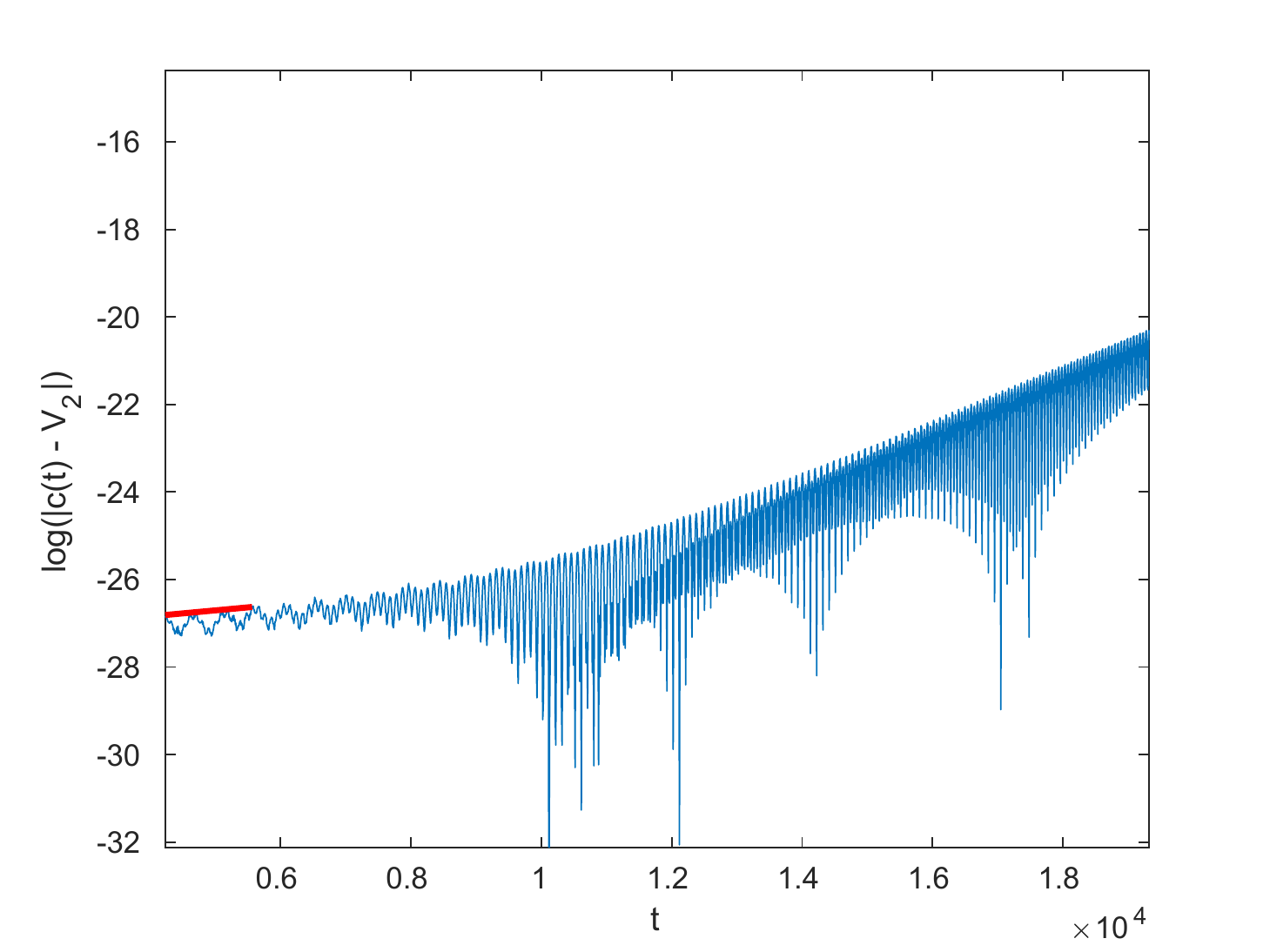}}
\subfloat[]
{\includegraphics[width=0.33\textwidth]{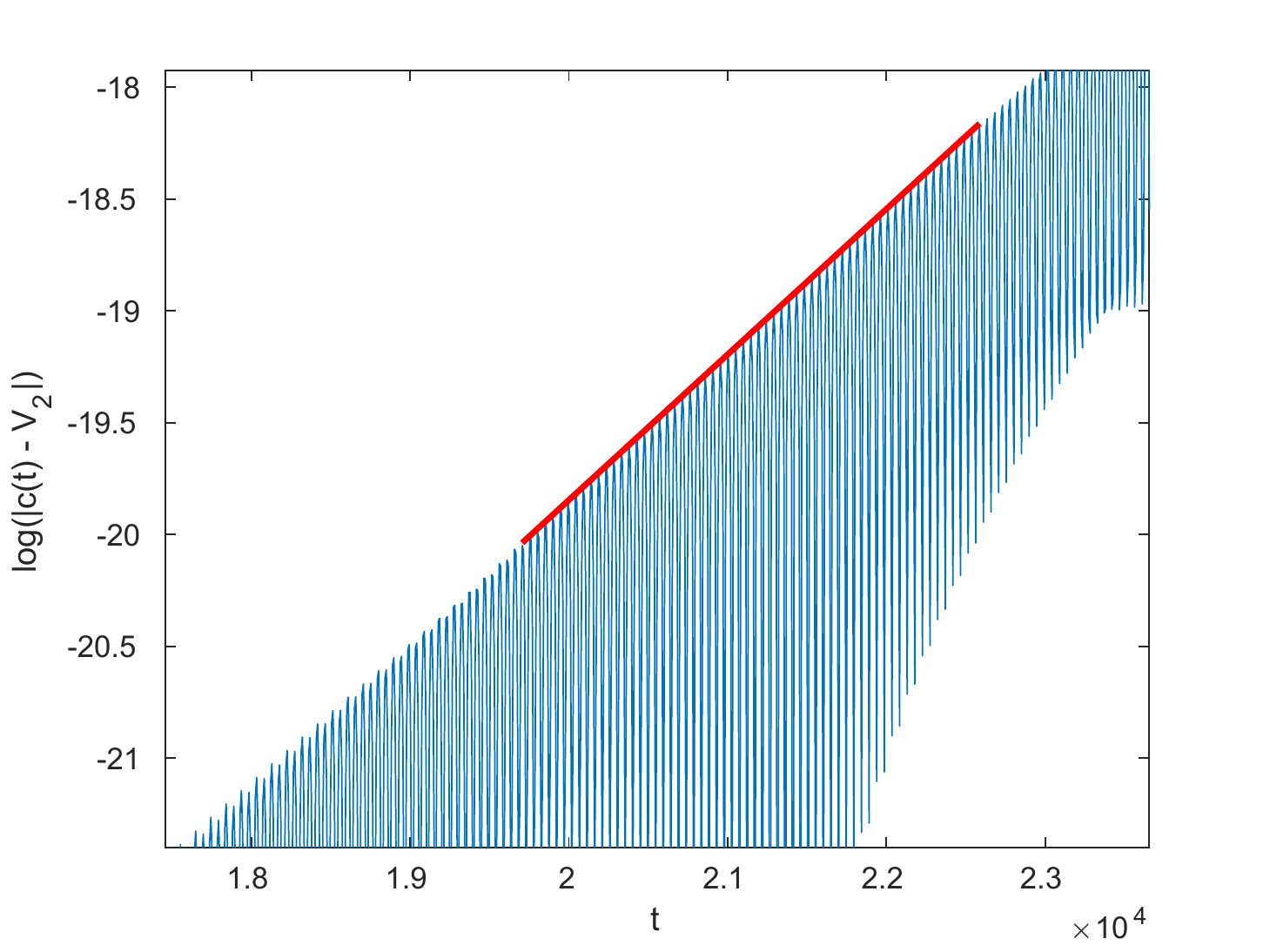}}
\caption{\footnotesize Time evolution of the absolute difference between the computed speed $c$ and the initial initial translational velocity $V_2 = 1/(3T)$ for the moving breather with the largest real Floquet multiplier $\mu = 1.0023$ perturbed along the corresponding unstable eigenmode. Panels (a) and (c) show the early and late stages of the evolution, while panel(b) depicts over the entire time span. The maximum modulus of the Floquet multipliers is $|\mu| = 1.0104$. In panels (a) and (c), the red lines correspond to the best linear fit measuring the growth rate of the wave. The line in panel (a) measures the initial growth due to the real instability, and the second line measures the growth due to the complex instability. In panel (b), the darker region corresponds to the emergence of the complex instability as the main factor in the growth of the perturbed moving breather. Here $N = 60$, $\omega = 2.355$, $V_1 = 1/3$, and the strength of the perturbation is $\epsilon = 10^{-7}$.
}
\label{fig:Res2355_instab}
\end{figure}
A key question is whether the real instability seen along the blue branch is a true instability given its relatively small size. To examine this, we perturbed a selected moving breather along its real unstable eigenmode with
a perturbation strength $\epsilon = 10^{-7}$, as explained in Sec.~\ref{sec:instab}. Here, $\mu = 1.0023$ is the largest real multiplier, $V_1 = 1/3$ and $\omega = 2.355$. In Fig.~\ref{fig:Res2355_instab}, we show a semilogarithmic plot of
the time evolution of the absolute difference of the computed velocity $c$ and the initial translational velocity $V_2 = 1/(3T)$. As can be seen, the growth of the perturbed moving breather has two regimes. The first, shown in Fig.~\ref{fig:Res2355_instab}(a), is dominated by the real instability associated with an eigenmode along which the dynamics was initially perturbed. The second, depicted in Fig.~\ref{fig:Res2355_instab}(c), is determined by the maximal-modulus Floquet multipliers $\mu = 0.5034 \pm 0.8761i$ with $|\mu| = 1.0104$.
The middle panel of Fig.~\ref{fig:Res2355_instab}(b) captures the transition from the former to the latter.
We note that this is different from the example shown in Sec.~\ref{sec:instab}, where the real Floquet multiplier corresponding to the eigenmode along which the moving breather was perturbed also had the largest modulus among the Floquet multipliers. The red lines in panels (a) and (c) measure the growth rate and have the slope $\ln(|\mu|)/(6T)$, where $\mu$ is the corresponding multiplier and we have used the fact that $V_1 = 1/3$ for the unperturbed breather. Comparing the lines of growth rate for maximum real multiplier and the complex multiplier with maximum modulus to the early and late stages, respectively, of the evolution in the simulation results yields an absolute difference of size $O(10^{-5})$ in both cases, indicating that the two regimes are indeed dominated by the two distinct types of instability. At  later times, the velocity evolution is similar to that for the examples discussed in Sec.~\ref{sec:instab}.
In short, this detailed examination of the associated dynamical evolution revealed that the instability growth
rates captured by our Floquet analysis, even when very small, accurately reflect the instability features of
the associated solutions and hence appear to be real features of the wave dynamics.

\bibliography{refer}
\end{document}